\documentclass[useAMS,usenatbib]{mn2e}
\usepackage{amsmath}
\usepackage{amsfonts}
\usepackage{bm}
\usepackage{booktabs}
\usepackage{natbib}
\usepackage{graphicx}
\usepackage{wasysym}
\usepackage[colorlinks=true,backref=page,bookmarksdepth=2]{hyperref}

\allowdisplaybreaks

\title[Restricted relativistic two-body problem with spin]
{
A first-order secular theory for the post-Newtonian two-body problem with spin -- I. The restricted case
}

\author[Francesco Biscani and Sante Carloni]{Francesco Biscani\thanks{E-mail:
bluescarni@gmail.com} and Sante Carloni\\
ESA -- Advanced Concepts Team, European Space Research Technology Center (ESTEC),\\
Keplerlaan 1, Postbus 299, 2200 AG Noordwijk The Netherlands
}

\begin{document}

\date{\today}

\pagerange{\pageref{firstpage}--\pageref{lastpage}}

\maketitle

\label{firstpage}

\begin{abstract}
We revisit the relativistic restricted two-body problem with spin employing a 
perturbation scheme based on Lie series. Starting from a post-Newtonian expansion of the field equations,
we develop a first-order secular theory that reproduces well-known relativistic effects such as
the precession of the pericentre and the Lense-Thirring and geodetic effects. Additionally,
our theory takes into full account the complex interplay between the various relativistic effects,
and provides a new explicit solution of the averaged equations of motion in terms of elliptic functions.
Our analysis reveals the presence of particular configurations for which
non-periodical behaviour can arise. The application of our results to real astrodynamical systems (such
as Mercury-like and pulsar planets) highlights the contribution of relativistic effects to the long-term evolution
of the spin and orbit of the secondary body.
\end{abstract}


The General Theory of Relativity (GR) has been one of the greatest achievements of the XX century. Its formulation has
revolutionised our way to understand the physical Universe and has led to the (sometimes unexpected) opening of entirely new research fields.
One striking example is the formulation of a consistent  theory describing the global behaviour of the Universe,
which was impossible in the framework of Newtonian mechanics \citep{peebles1993principles}.

In Einstein's theory the spacetime is thought as a four dimensional pseudo-Riemannian  manifold whose geometry interacts non-linearly with matter.
Because of this feature, in general calculations in GR require much more work than the corresponding ones in Newtonian
gravity and this fact has made more difficult the understanding of the physics of Einsteinian gravity.

Another consequence of this difficulty falls in the context of testing Einstein's gravity: our inability to solve in general the gravitational
field equations limits the possibility of devising full tests of GR. Instead, most of the investigation on the validity of this theory
was performed in a regime of weak field leaving the strong field regime almost completely unexplored\footnote{There are in truth other reasons why all our
tests are focused on weak field: we live in a weak gravitational field and it is impossible to control gravitational fields
like we do, for example, with electromagnetism.}. In fact, all the tests of GR performed up to now belong to the class of weak field experiments,
from the so-called ``classical tests'' of Relativity \citep{peebles1993principles} to the latest tests based, e.g., on the motion of
interplanetary probes \citep{Hees:2011wt}.

Einstein himself was well aware of the fact that the understanding
of his theory in the perturbative regime was crucial for its testing
and its application to the problems in which Newtonian gravity was
most successful. For this reason in the 1930s he developed a perturbative approach
able to describe the subtle changes induced on the Newtonian evolution
of the bodies in the Solar System by General Relativity. The so-called
Einstein-Infeld-Hoffman (EIH) equations are the results at first post-Newtonian level of this attempt \citep{Einstein:1938kq}\footnote{To be precise,
in the same period other researchers proposed alternative methods based on different assumptions
\citep{fock-book,papapetrou1974lectures}, but they all lead to the same results.}.  The derivation of these equations was the the birth of a new field
called ``Relativistic Celestial Mechanics'' (RCM) \citep{brumberg1991essential,kopeikin2011relativistic}.

The research in RCM has not stopped since then. The post-Newtonian equations
have been analysed in many different ways and new and more powerful
approaches to RCM have been proposed \citep[see][]{1991PhRvD..43.3273D,1993PhRvD..47.3124D,1994PhRvD..49..618D,Damour:227658}.
In spite of these advances, the resolution of the post-Newtonian equations still remains a formidable task.
In particular, the interaction between orbital and rotational motion has not been fully solved analytically to this date.

In this paper we will focus on the post-Newtonian equations for a restricted two spinning bodies system at the 1PN level.
Our target is to obtain a complete description of the leading correction of GR on Newtonian mechanics with a specific focus on the spin interaction effects.
Such target will be accomplished using modern perturbative methods based on Lie series \citep{hori_theory_1966,deprit_canonical_1969}. This approach allows us
to derive a fully-analytical description of the long-term dynamical evolution of the system using semi-automatised computer algebra. Advantages
of the the Lie series procedure over traditional perturbation techniques include the availability of explicit formul\ae{} to convert between the averaged
and non-averaged orbital elements, and the possibility to straightforwardly extend the treatment to higher post-Newtonian orders.
Using the Lie series technique, we will re-derive the standard precession effects typical
of the restricted two-body problem in General Relativity, but we will also give new analytical relations describing the evolution of spin and
orbital angular momentum of the secondary body. In fact, we will be able to explore the full evolution of the orbital parameters of the secondary
body and to quantify the effect of the spin-orbit and spin-spin interactions.

The use of the Lie series technique in RCM is relatively sparse, and, as far as we have been able to verify, it constitutes by itself a novelty in the analysis
of relativistic spin-spin and spin-orbit interactions\footnote{
We were able to locate two papers in the existing literature that tackle the solution of the post-Newtonian equations using Lie series \citep{1988CeMec..43..193R,
heimberger_relativistic_1990}, none of them considering spin interactions.}. In this first paper we have thus chosen to focus
on the restricted problem in order to lay the foundations of our method in the simplest case of interest. In subsequent publications,
we will tackle the full post-Newtonian two-body problem with spin building on the formalism introduced in the present paper.

% The paper is organized in the following way. In Section \ref{sec:PPN_EIH} we will sketch briefly the derivation of the EIH equations. In Section \ref{sec:lag_ham}
% we will specialize these equations to the case of two rotating bodies and we will recast them in Hamiltonian form,
% which will be the starting point of the Lie series approach. In Section \ref{sec:canonical_vars} we will introduce some useful canonical variables:
% the Serret-Andoyer and Delaunay variables, and we will use them to rewrite the Hamiltonian.
% In Section \ref{sec:lie_series} we will briefly review the Lie series approach and we will apply it to the Hamiltonian obtained in Section \ref{sec:lag_ham}.
% In Section \ref{sec:particular_cases} we will consider some particular cases obtaining the formula\ae{} for the perihelion shift,
% the Lens-Thirring Effect and the de Sitter Precession. In Section \ref{sec:general_case} we will then consider the general case, providing also
% a phase space analysis of the two body system. In Section \ref{sec:physical_systems} we provide some examples inspired by physical systems. Finally Section
% \ref{sec:conclusions} is dedicated to the conclusions.

\section{Hamiltonian formulation}
The starting point of our derivation is the well-known 1PN Hamiltonian of the two-body problem with spin, which, after  reduction to the centre-of-mass
frame, reads \citep{Barker:1970vv,BARKER:1979wb,damour_coalescence_2001}
\begin{equation}
\mathcal{H} = \mathcal{H}_\textnormal{N} + \epsilon\mathcal{H}_1.
\label{eq:Ham_00}
\end{equation}
Here $\epsilon = 1 / c^2$ is chosen as the ``smallness parameter'' of our perturbation theory and
\begin{equation}
\mathcal{H}_\textnormal{N} = \frac{1}{2}\frac{\bm{J}_1^2}{I_1} + \frac{1}{2}\frac{\bm{J}_2^2}{I_2} + \frac{\bm{p}^2}{2\mu}-\frac{\mathcal{G}M\mu}{r}
\label{eq:Ham_00_N}
\end{equation}
is the Newtonian Hamiltonian (representing the unperturbed problem). $\mathcal{H}_1$ expands as
\begin{equation}
\mathcal{H}_1 = \mathcal{H}_\textnormal{PN}+\mathcal{H}_\textnormal{SO}+\mathcal{H}_\textnormal{SS},
\label{eq:Ham_split_00}
\end{equation}
where
\begin{align}
\mathcal{H}_\textnormal{PN} & = \mu\left\{ \frac{1}{8}\left( 3\nu-1\right)\frac{\bm{p}^4}{\mu^4} 
\vphantom{\left. -\frac{\mathcal{G}M}{2r}\left[\left(3+\nu\right)\frac{\bm{p}^2}{\mu^2}+\nu\left(\bm{n}\cdot\frac{\bm{p}}{\mu}\right)^2\right] +\frac{\mathcal{G}^2M^2}{2r^2} \right\}}
\right.\notag\\
&\quad \left. -\frac{\mathcal{G}M}{2r}\left[\left(3+\nu\right)\frac{\bm{p}^2}{\mu^2}+\nu\left(\bm{n}\cdot\frac{\bm{p}}{\mu}\right)^2\right] +\frac{\mathcal{G}^2M^2}{2r^2} \right\}
\label{eq:Ham_00_PN}
\end{align}
is the post-Newtonian \emph{orbital} Hamiltonian,
\begin{equation}
\mathcal{H}_\textnormal{SO} = \frac{2\mathcal{G}}{r^3}\left[\left(1+\frac{3}{4}\frac{m_2}{m_1} \right)\bm{J}_1 +
\left(1+\frac{3}{4}\frac{m_1}{m_2} \right)\bm{J}_2\right]\cdot\left(\bm{r}\times\bm{p}\right)
\label{eq:Ham_00_SO}
\end{equation}
represents the spin-orbit interaction, and
\begin{equation}
\mathcal{H}_\textnormal{SS} = \frac{\mathcal{G}}{r^3}\left[3\left(\bm{J}_1\cdot\bm{n}\right)\left(\bm{J}_2\cdot\bm{n}\right)-\bm{J}_1\cdot\bm{J}_2\right]
\label{eq:Ham_00_SS}
\end{equation}
represents the spin-spin coupling. In these formul\ae{}, $\mathcal{G}$ is the universal gravitational constant,
$m_1$ and $m_2$ are the masses of the two bodies, $M=m_1+m_2$, $\mu=m_1m_2/M$ is the reduced mass, $\nu = \mu / M$,
$\bm{p}=\bm{p}_1=-\bm{p}_2$ is the canonical momentum of body 1, and $\bm{r}=r\bm{n}$ is the vector connecting body 2 to body 1.
The structure of this Hamiltonian and, above all, the nature of the spin
vectors, have been a matter of debate in the past few years \citep[see][]{damour_effective_2008}.
More recently, \citet{wu_symplectic_2010} have proposed a symplectic formulation of the spin vectors in terms of
cylindrical-like coordinates.
Here, we follow the interpretation of \citet{barker_gravitational_1975,barker_lagrangian-hamiltonian_1976,BARKER:1979wb}
and \citet{wex_second_1995} in identifying the spins $\bm{J}_i$ as the rotational angular
momenta of spherical rigid bodies, so that
\begin{equation}
\bm{J}_i = I_i\bm{\omega}_i,
\end{equation}
where $I_i$ is the moment of inertia and $\bm{\omega}_i$ the rotational angular velocity vector of body $i$.

In the present formulation of the Hamiltonian, we have
dropped from $\mathcal{H}_{\textnormal{SS}}$ the spin-quadratic contributions due to monopole-quadrupole interaction\footnote{
The monopole-quadrupole interaction terms for compact bodies will depend on their internal structure and equations of state.
For black holes, they are given, e.g., in \citep{damour_coalescence_2001}.
} to focus on the leading relativistic effects. Besides, as pointed out by \citet{damour_coalescence_2001}, at this PN order
we can expect to obtain physically reliable results in case of moderate spins and orbital velocities.

The equations of motion generated by \eqref{eq:Ham_00} can be obtained via the familiar Hamiltonian canonical equations. The orbital momenta
and coordinates are represented respectively by $\bm{p}$ and $\bm{r}$; the spin coordinates and momenta are represented by the Euler angles
and their conjugate generalised momenta in terms of which the components of the angular velocities $\bm{\omega}_i$ are expressed. At this stage,
we are not concerned with the exact functional form of such dependence, as we will introduce a more convenient set of variables in \S \ref{sec:aa_vars}.
We refer to \citet{gurfil_serret-andoyer_2007} for a derivation of the Hamiltonian formulation of rigid-body dynamics.

The reduction of Hamiltonians \eqref{eq:Ham_00_N}-\eqref{eq:Ham_00_SS} to the restricted case in which $m_2 \gg m_1$ and $\left|\bm{J}_2\right| \gg \left|\bm{J}_1\right|$
reads
\begin{equation}
\begin{split}
\mathcal{H}_\textnormal{N} & = \frac{1}{2}\frac{\bm{J}_1^2}{I_1} + \frac{\bm{p}_1^2}{2m_1}-\frac{\mathcal{G}m_1m_2}{r},\\
\mathcal{H}_\textnormal{PN} & = m_1\left( -\frac{1}{8}\frac{\bm{p}_1^4}{m_1^4} -
\frac{3}{2}\frac{\mathcal{G}m_2}{r}\frac{\bm{p}_1^2}{m_1^2} +\frac{\mathcal{G}^2m_2^2}{2r^2} \right), \\
\mathcal{H}_\textnormal{SO} & = \frac{2\mathcal{G}}{r^3}\left(\frac{3}{4}\frac{m_2}{m_1}\bm{J}_1 +
\bm{J}_2\right)\cdot\left(\bm{r}\times\bm{p}_1\right),\\
\mathcal{H}_\textnormal{SS} & = \frac{\mathcal{G}}{r^3}\left[3\left(\bm{J}_1\cdot\bm{n}\right)\left(\bm{J}_2\cdot\bm{n}\right)-\bm{J}_1\cdot\bm{J}_2\right],
\label{eq:Ham_restr_00}
\end{split}
\end{equation}
where $\bm{J}_2$ is now considered as a constant of motion that can be dropped from $\mathcal{H}_\textnormal{N}$. Without loss of generality, we can
orient the reference system (now centred on body 2) in such a way that the constant $\bm{J}_2$ is aligned to the positive $z$ axis, so that
\begin{equation}
\bm{J}_2 = \left(0,0,J_2\right),
\end{equation}
with $J_2=\left| \bm{J}_2 \right|$.
\section{Action-angle variables}
\label{sec:aa_vars}
In preparation for the application of the Lie series peturbative approach
in the study of Hamiltonian \eqref{eq:Ham_restr_00},
we want to
express the Hamiltonian in a coordinate system of action-angle variables
\citep{arnold_mathematical_1989} for the
unperturbed problem. For the orbital variables, a common choice is the set
of Delaunay elements \citep{morbidelli_modern_2002}. For the rotational motion,
a suitable choice in our case is that of Serret-Andoyer (SA)
variables \citep{gurfil_serret-andoyer_2007}. Both sets of variables are introduced
formally as canonical transformations.

Before proceeding, we need to discuss briefly the nature of the canonical momenta
appearing in Hamiltonian \eqref{eq:Ham_restr_00}. Post-Newtonian Hamiltonians are derived
from Lagrangians in which the generalised velocities appear not only in the kinetic
terms of the Newtonian portion of the Lagrangian, but also in the relativistic perturbation
(see, e.g., \citet{landau1975classical} and \citet{straumann_general_1984}). As a
consequence, after switching to the Hamiltonian formulation via the usual Legendre
transformation procedure, the post-Newtonian canonical Hamiltonian momenta will
differ from the Newtonian ones by terms of order $1/c^2$. This discrepancy will
carry over to any subsequent canonical transformation, including the introduction
of Delaunay and SA elements. \citet{1988CeMec..43..193R} and
\citet{heimberger_relativistic_1990} analyse in detail the connection between Newtonian
and post-Newtonian Delaunay orbital elements.

However, in the present work we are concerned with the secular variations of orbital
and rotational elements, for which the discrepancy discussed above is of little consequence.
For if a secular motion (e.g., a precession of the node) results from our analysis
in terms of post-Newtonian elements, it will affect within an accuracy of $1/c^2$ also
the Newtonian elements\footnote{Vice versa, short-term periodic variations of the
post-Newtonian elements will have
amplitudes of order $1/c^2$ and the precise connection with the Newtonian elements
therefore becomes important.}. Indeed, we will show how our analysis in terms of post-Newtonian elements reproduces exactly the formul\ae{} of known secular relativistic effects.
\subsection{Delaunay elements}
The Delaunay arguments $\left(L,G,H,l,g,h\right)$ can be introduced via the following
standard relations
\citep{morbidelli_modern_2002}:
\begin{equation}
\begin{aligned}L & =\sqrt{\mathcal{G}m_{2}a}, & l & =M,\\
G & =L\sqrt{1-e^{2}}, & g & =\omega,\\
H & =G\cos i, & h & =\Omega.
\end{aligned}
\label{eq:Del_00}
\end{equation}
Here $a$, $e$, $i$, $M$, $\omega$ and $\Omega$ are the classical
Keplerian orbital elements describing the trajectory of the secondary body $m_{1}$
around the primary $m_{2}$. The Keplerian elements are in turn
related to the cartesian orbital momentum $\bm{p}_1$ and position $\bm{r}$
via well-known relations
(e.g., see \citet{murray_solar_2000}). In eqs. \eqref{eq:Del_00}, $\left(L,G,H\right)$
play the role of generalised momenta, $\left(l,g,h\right)$ are the generalised coordinates.
\subsection{Serret-Andoyer variables}
The Serret-Andoyer (SA) variables describe the rotational motion of a rigid body in terms of
orientation angles and rotational angular momentum.
In our specific case, the bodies are perfectly
spherical and thus the introduction of SA elements
is simplified with respect to the general case. We refer to
\citet{gurfil_serret-andoyer_2007} for an exhaustive treatment of the SA formalism in the
general case.

We start by expressing $\bm{J}_1$ in a coordinate system where
$\bm{J}_1$ itself is aligned to the $z$ axis, so that
$\bm{J}_1=\left(0,0,\left|\bm{J}_{1}\right|\right)$\footnote{
This reference system is referred to as ``invariable''
in \citet{gurfil_serret-andoyer_2007}.}. In order to
express the components of $\bm{J}_1$ in the centre-of-mass reference system, we compose
two rotations: the first one around the $x$ axis by $-I$ (the inclination
of $\bm{J}_1$), the second one around the new $z$ axis by $-\tilde{h}$ (the
nodal angle of $\bm{J}_1$, analogous to the $h$ angle in the
Delaunay elements set or to the longitude of the node in
the set of Keplerian elements), resulting in the composite rotation
matrix
\begin{equation}
\left[\begin{array}{ccc}
\cos\tilde{h} & -\sin\tilde{h}\cos I & \sin I\sin\tilde{h}\\
\sin\tilde{h} & \cos I\cos\tilde{h} & -\sin I\cos\tilde{h}\\
0 & \sin I & \cos I
\end{array}\right].
\end{equation}
In the SA formalism, $\cos I=J_{1,z} / \left|\bm{J}_1\right|$ (where
$J_{1,z}$ is the $z$ component of the rotational angular momentum
in the centre-of-mass reference system). Thus, in the centre-of-mass reference system,
the components
of $\bm{J}_1$ read
\begin{equation}
\bm{J}_1=\left(\sqrt{\left|\bm{J}_1\right|^2-J_{1,z}^2}\sin\tilde{h},
-\sqrt{\left|\bm{J}_1\right|^2-J_{1,z}^2}\cos\tilde{h},J_{1,z}\right).
\end{equation}
In the SA variables set, $\left(\left|\bm{J}_1\right|,J_{1,z}\right)$ are the generalised
momenta, and $\left(\tilde{g},\tilde{h}\right)$ are the conjugate coordinates. Note that,
due to the spherical symmetry of the bodies, the angle $\tilde{g}$ conjugated
to the action $\left|\bm{J}_1\right|$ does not appear
in the expression of the Hamiltonian and
$\left|\bm{J}_1\right|$ is thus conserved.
\subsection{Final form of the Hamiltonian}
We are now ready to express the Hamiltonians \eqref{eq:Ham_restr_00} in terms of Delaunay and Serret-Andoyer
arguments. Before proceeding to the substitutions, in order to simplify the notation
we rescale the Hamiltonian via an extended canonical transformation, dividing the
Hamiltonian and the generalised momenta by $m_1$. The resulting expressions for
$\mathcal{H}_\textnormal{N}$ and $\mathcal{H}_1$ in terms of the momenta
$\left(L,G,H,\tilde{G},\tilde{H}\right)$ and coordinates $\left(l,g,h,\tilde{g},\tilde{h}\right)$
read
\begin{align}
\mathcal{H}_{\textnormal{N}}& =
\frac{1}{2}\mathcal{I}_{1}\tilde{G}^{2}-\frac{\mathcal{G}^{2}m_{2}^{2}}{2L^{2}},\label{eq:Ham_F_N}\\
\mathcal{H}_{1} & =-\frac{1}{8}\frac{\mathcal{G}^{4}m_{2}^{4}}{L^{4}}+\frac{1}{r}\frac{2\mathcal{G}^{3}m_{2}^{3}}{L^{2}}-\frac{1}{r^{2}}3\mathcal{G}^{2}m_{2}^{2}\nonumber \\
 & \quad+\frac{\mathcal{G}}{r^{3}}\left\{2J_{2}H+\frac{3J_{2}G_{xy}^{2}\tilde{H}}{2G^{2}}+\frac{3m_{2}\tilde{H}H}{2}-J_{2}\tilde{H}\right.\nonumber \\
 & \quad+\left(\frac{3m_{2}}{2}\tilde{G}_{xy}G_{xy}-\frac{3}{2}J_{2}\frac{HG_{xy}\tilde{G}_{xy}}{G^{2}}\right)\cos\left(\tilde{h}-h\right)\nonumber \\
 & \quad+3J_{2}\left[-\frac{1}{2}\frac{G_{xy}^{2}\tilde{H}}{G^{2}}\cos\left(2f+2g\right)\right.\nonumber \\
 & \quad-\frac{1}{4}\frac{G_{xy}\tilde{G}_{xy}}{G}\left(1-\frac{H}{G}\right)\cos\left(2f+2g+\tilde{h}-h\right)\nonumber \\
 & \quad\left.\left.\vphantom{\frac{3J_{2}G_{xy}^{2}\tilde{H}}{2G^{2}}}
 +\frac{1}{4}\frac{G_{xy}\tilde{G}_{xy}}{G}\left(1+\frac{H}{G}\right)\cos\left(2f+2g-\tilde{h}+h\right)\right]\right\}.\label{eq:Ham_F_1}
\end{align}
For convenience, we have summarised in Table \ref{tab:Ham_F_vars} the meaning of the symbolic variables appearing in Hamiltonians \eqref{eq:Ham_F_N}-\eqref{eq:Ham_F_1}.
We have to stress how in these expressions, $r$, $G_{xy}$, $\tilde{G}_{xy}$ and $f$ have to be regarded as implicit functions of the canonical momenta
and coordinates. Specifically,
\begin{align}
r & = r\left(L,G,l\right),\\
G_{xy} & = G_{xy}\left(G,H\right),\\
\tilde{G}_{xy} & = \tilde{G}_{xy}\left(\tilde{G},\tilde{H}\right),\\
f & = f\left(L,G,l\right).
\end{align}
We are now ready to analyse the Hamiltonian using the Lie series method.
\begin{table*}
\caption{Summary and explanation of the symbolic variables appearing in Hamiltonians \eqref{eq:Ham_F_N}-\eqref{eq:Ham_F_1}.}
\label{tab:Ham_F_vars}
\centering
\begin{tabular}{cll}
\toprule
Symbol & Meaning & Alternate expression\\
\midrule
$L$ & normalised square root of the semi-major axis & $\sqrt{\mathcal{G}m_{2}a}$\\
$G$ & norm of the orbital angular momentum $\bm{r}\times\bm{p}_1$ normalised by $m_1$ & $L\sqrt{1-e^2}$\\
$H$ & $z$ component of the orbital angular momentum normalised by $m_1$ & $G\cos{i}$\\
$\tilde{G}$ & norm of spin $\bm{J}_1$ normalised by $m_1$ & $\left|\bm{J}_1\right|/m_1$\\
$\tilde{H}$ & $z$ component of spin $\bm{J}_1$ normalised by $m_1$ & $\tilde{G}\cos{I}$\\
$l$ & mean anomaly & $M$\\
$g$ & argument of pericentre & $\omega$\\
$h$ & longitude of the ascending node & $\Omega$\\
$\tilde{h}$ & nodal angle of spin $\bm{J}_1$ & -\\
\midrule
$f$& true anomaly& -\\
$\mathcal{I}_1$& inverse of the moment of inertia of body 1 normalised by $m_1$& $m_1/I_1$\\
$G_{xy}$& norm of the projection of $\bm{r}\times\bm{p}_1$ on the $xy$ plane normalised by $m_1$& $\sqrt{G^2-H^2}$\\
$\tilde{G}_{xy}$& norm of the projection of spin $\bm{J}_1$ on the $xy$ plane normalised by $m_1$& $\sqrt{\tilde{G}^2-\tilde{H}^2}$\\
$J_2$ & norm of spin $\bm{J}_2$ & $\left|\bm{J}_2\right|$\\
$r$& distance between body 1 and body 2& -\\
$m_1$, $m_2$ & masses of the two bodies & -\\
$\mathcal{G}$ & universal gravitational constant & -\\
\bottomrule
\end{tabular}
\end{table*}
\section{Perturbation theory via Lie series}
The Lie series perturbative methodology \citep{hori_theory_1966,deprit_canonical_1969}
aims at simplifying the Hamiltonian of the problem via a quasi-identity canonical transformation of coordinates
depending on a generating function $\chi$ to be determined. The Lie series transformation reads
\begin{equation}
\mathcal{H}^{\prime}=\mathcal{S}_{\chi}^{\epsilon}\mathcal{H}=\sum_{n=0}^\infty\frac{\epsilon^{n}}{n!}\mathcal{L}_{\chi}^{n}\mathcal{H},
\end{equation}
where $\mathcal{L}_{\chi}^{n}$ is the
Lie derivative of $n$-th order with generator $\chi$, $\mathcal{H}^{\prime}$ is the transformed Hamiltonian,
and $\mathcal{H}$ is the original Hamiltonian in which the new momenta and coordinates have been formally substituted.
At the first order in $\epsilon$
the Lie derivative degenerates to a Poisson bracket, and the transformation becomes
\begin{equation}
\mathcal{H}^{\prime}=\mathcal{H}_\textnormal{N}+\epsilon\underbrace{\left(\left\{ \mathcal{H}_\textnormal{N},\chi\right\}
+\mathcal{H}_{1}\right)}_{\mathcal{K}}+\textnormal{O}\left(\epsilon^2\right).
\label{eq:lie_S_prime}
\end{equation}
We need then to solve the homological equation \citep{arnold_mathematical_1989}
\begin{equation}
\left\{ \mathcal{H}_\textnormal{N},\chi\right\} +\mathcal{H}_{1}=\mathcal{K},
\label{eq:h_eq_00}
\end{equation}
where $\chi$ and $\mathcal{K}$ (the new perturbed Hamiltonian resulting from the transformation of coordinates) have to be determined
with the goal of obtaining some form of simplification in $\mathcal{K}$.
Since the unperturbed Hamiltonian depends only on the two actions $L$
and $\tilde{G}$, we have
\begin{equation}
\left\{ \mathcal{H}_\textnormal{N},\chi\right\} =-\frac{\partial\mathcal{H}_\textnormal{N}}{\partial L}\frac{\partial\chi}{\partial l}
=-\frac{\mathcal{G}^{2}m_{2}^{2}}{L^{3}}\frac{\partial\chi}{\partial l},
\end{equation} 
where the partial derivative of $\chi$ with respect to $\tilde{g}$ (the coordinate conjugated to $\tilde{G}$)
can be set to zero as $\tilde{G}$ is
a constant of motion. The homological equation \eqref{eq:h_eq_00} then reads
\begin{equation}
\chi=\int\frac{L^{3}}{\mathcal{G}^{2}m_{2}^{2}}\left(\mathcal{H}_{1}-\mathcal{K}\right)dl.
\label{eq:chi_integral}
\end{equation}
The technical aspects of the solution of this integral, including the choice of $\mathcal{K}$, are detailed in Appendix \ref{sec:closed_form_av}. Here
we will limit ourselves to the following considerations:
\begin{itemize}
\item the integral in eq. \eqref{eq:chi_integral} essentially represents an averaging over the mean motion $l$.
Consequently, the new momenta and coordinates generated by the Lie series transformation are the \emph{mean}
counterparts of the original momenta and coordinates;
\item thanks to the functional form of Hamiltonians \eqref{eq:Ham_F_N}-\eqref{eq:Ham_F_1}, the averaging procedure removes at the same time
both $l$ and $g$ from the averaged Hamiltonian;
\item the integration is performed in closed form, that is, without resorting to Fourier-Taylor expansions in terms of
mean anomaly and eccentricity \citep{deprit_delaunay_1982,palacian_closed-form_2002}.
The results are thus valid also for highly-eccentric orbits.
\end{itemize}
The computations involved in the averaging procedure have been carried out with the Piranha computer algebra system \citep{biscani_design_2008}.
As usual when operating with Lie series transformations,
from now on we will refer to the mean momenta and coordinates
$\left(L^\prime,G^\prime,H^\prime,\tilde{G}^\prime,\tilde{H}^\prime\right)$ and $\left(l^\prime,g^\prime,h^\prime,\tilde{g}^\prime,\tilde{h}^\prime\right)$
with their original names $\left(L,G,H,\tilde{G},\tilde{H}\right)$ and $\left(l,g,h,\tilde{g},\tilde{h}\right)$, in order to simplify the notation.

After having determined $\chi$ from eq. \eqref{eq:chi_integral}, the averaged Hamiltonian generated by eq. \eqref{eq:h_eq_00} reads,
in terms of mean elements,
\begin{equation}
\mathcal{H}^{\prime}=\mathcal{H}_\textnormal{N}+\epsilon\left[\mathcal{E}_{0}+\mathcal{E}_{1}\cos\left(\tilde{h}-h\right)\right],
\label{eq:H_av_EE}
\end{equation}
where $\mathcal{E}_{0}$ and $\mathcal{E}_{1}$ are functions of the mean momenta only.
% \begin{align}
% \mathcal{E}_{0} & = \frac{1}{2}\frac{{J_2}{\mathcal{G}}^{4}{\tilde{H}}{m_2}^{3}}{{G}^{3}{L}^{3}}
% +\frac{15}{8}\frac{{\mathcal{G}}^{4}{m_2}^{4}}{{L}^{4}}
% +\frac{3}{2}\frac{{H}{\mathcal{G}}^{4}{\tilde{H}}{m_2}^{4}}{{G}^{3}{L}^{3}}\notag\\
% &\quad +2\frac{{H}{J_2}{\mathcal{G}}^{4}{m_2}^{3}}{{G}^{3}{L}^{3}}
% -\frac{3}{2}\frac{{H}^{2}{J_2}{\mathcal{G}}^{4}{\tilde{H}}{m_2}^{3}}{{G}^{5}{L}^{3}}-3\frac{{\mathcal{G}}^{4}{m_2}^{4}}{{G}{L}^{3}},\\
% \mathcal{E}_{1} & = -\frac{3}{2}\frac{{G_{xy}}{H}{J_2}{\mathcal{G}}^{4}{\tilde{G}_{xy}}{m_2}^{3}}{{G}^{5}{L}^{3}}
% +\frac{3}{2}\frac{{G_{xy}}{\mathcal{G}}^{4}{\tilde{G}_{xy}}{m_2}^{4}}{{G}^{3}{L}^{3}}.
% \end{align}
In order to further reduce the number of degrees of freedom, we perform a final canonical transformation\footnote{
This transformation is reminiscent of the treatment of single-resonance dynamics in the N-body problem (see Lemma 
4 in \citet{morbidelli_quantitative_1993}).
} that compresses the
coordinates $h$ and $\tilde{h}$ in a single coordinate $h_{\ast}$ (replacing $h$) and introduces a new momentum $\tilde{H}_{\ast}$
(replacing $\tilde{H}$) via the relations
\begin{align}
\tilde{H}_{\ast} & =H+\tilde{H},\label{eq:Hts}\\
h_{\ast} & =h-\tilde{h}.
\label{eq:hs}
\end{align}
After this transformation, the final averaged Hamiltonian reads
\begin{equation}
\mathcal{H}^{\prime}=\mathcal{H}_\textnormal{N}+\epsilon\left(\mathcal{F}_{0}+\mathcal{F}_{1}\cos h_\ast \right),
\label{eq:Ham_av_00}
\end{equation}
where $\mathcal{F}_{0}$ and $\mathcal{F}_{1}$ are functions of the mean momenta only,
\begin{align}
\mathcal{F}_{0} & = \frac{1}{2}\frac{{J_2}{\mathcal{G}}^{4}{\tilde{H}_\ast}{m_2}^{3}}{{G}^{3}{L}^{3}}
+\frac{3}{2}\frac{{H}^{3}{J_2}{\mathcal{G}}^{4}{m_2}^{3}}{{G}^{5}{L}^{3}}+\frac{15}{8}\frac{{\mathcal{G}}^{4}{m_2}^{4}}{{L}^{4}}\notag\\
&\quad +\frac{3}{2}\frac{{H}{J_2}{\mathcal{G}}^{4}{m_2}^{3}}{{G}^{3}{L}^{3}}-\frac{3}{2}\frac{{H}^{2}{J_2}{\mathcal{G}}^{4}{\tilde{H}_\ast}{m_2}^{3}}{{G}^{5}{L}^{3}}
-\frac{3}{2}\frac{{H}^{2}{\mathcal{G}}^{4}{m_2}^{4}}{{G}^{3}{L}^{3}}\notag\\
&\quad +\frac{3}{2}\frac{{H}{\mathcal{G}}^{4}{\tilde{H}_\ast}{m_2}^{4}}{{G}^{3}{L}^{3}}-3\frac{{\mathcal{G}}^{4}{m_2}^{4}}{{G}{L}^{3}},\label{eq:F0}\\
\mathcal{F}_{1} & = -\frac{3}{2}\frac{{G_{xy}}{H}{J_2}{\mathcal{G}}^{4}{\tilde{G}_{xy\ast}}{m_2}^{3}}{{G}^{5}{L}^{3}}+\frac{3}{2}\frac{{G_{xy}}{\mathcal{G}}^{4}{\tilde{G}_{xy\ast}}{m_2}^{4}}{{G}^{3}{L}^{3}},
\label{eq:F1}
\end{align}
and $\tilde{G}_{xy\ast}$ is the mean $\tilde{G}_{xy}$ (see Table \ref{tab:Ham_F_vars}) expressed in terms of the new mean momentum $\tilde{H}_{\ast}$:
\begin{equation}
\tilde{G}_{xy\ast}=\sqrt{\tilde{G}^{2}-\left(\tilde{H}_{\ast}-H\right)^{2}}.
\end{equation}
We are now ready to proceed to an analysis of the averaged Hamiltonian \eqref{eq:Ham_av_00}.
\section{Analysis of the averaged Hamiltonian}
\subsection*{Preliminary considerations}
The one degree-of-freedom averaged Hamiltonian \eqref{eq:Ham_av_00} is expressed in terms of the mean momenta
$\left(L,G,H,\tilde{G},\tilde{H}_\ast\right)$ and conjugate mean coordinates
$\left(l,g,h_\ast,\tilde{g},\tilde{h}\right)$. The equations of motion
generated by Hamiltonian \eqref{eq:Ham_av_00} read:
\begin{equation}
\begin{aligned}\frac{dL}{dt} & =0,\\
\frac{dG}{dt} & =0,\\
\frac{dH}{dt} & =\epsilon\mathcal{F}_{1}\sin h_{\ast},\\
\frac{d\tilde{G}}{dt} & =0,\\
\frac{d\tilde{H}_{\ast}}{dt} & =0,
\end{aligned}
\label{eq:averaged_momenta}
\end{equation}
and
\begin{equation}
\begin{aligned}\frac{dl}{dt} & =\frac{\mathcal{G}^{2}m_{2}^{2}}{L^{3}}+\epsilon\left(\frac{\partial\mathcal{F}_{0}}{\partial L}+\frac{\partial\mathcal{F}_{1}}{\partial L}\cos h_{\ast}\right),\\
\frac{dg}{dt} & =\epsilon\left(\frac{\partial\mathcal{F}_{0}}{\partial G}+\frac{\partial\mathcal{F}_{1}}{\partial G}\cos h_{\ast}\right),\\
\frac{dh_{\ast}}{dt} & =\epsilon\left(\frac{\partial\mathcal{F}_{0}}{\partial H}+\frac{\partial\mathcal{F}_{1}}{\partial H}\cos h_{\ast}\right),\\
\frac{d\tilde{g}}{dt} & =\mathcal{I}_{1}\tilde{G}+\epsilon\frac{\partial\mathcal{F}_{1}}{\partial\tilde{G}}\cos h_{\ast},\\
\frac{d\tilde{h}}{dt} & =\epsilon\left(\frac{\partial\mathcal{F}_{0}}{\partial\tilde{H}_{\ast}}+\frac{\partial\mathcal{F}_{1}}{\partial\tilde{H}_{\ast}}\cos h_{\ast}\right).
\end{aligned}
\label{eq:averaged_angles}
\end{equation}
A first straightforward consequence of the form of the averaged
Hamiltonian is the conservation of all mean momenta apart from $H$. In particular, the conservation
of the mean momentum $\tilde{H}_\ast=H+\tilde{H}$ corresponds to the
conservation of the $z$ component of the total mean angular momentum of the system.
We refer the reader to Appendix \ref{sec:full_form} for the full functional form of the partial derivatives
of $\mathcal{F}_0$ and $\mathcal{F}_1$ with respect to the mean momenta.

We are now going to proceed to the derivation of well-known relativistic effect
in a series of simplified cases of Hamiltonian \eqref{eq:Ham_av_00}.
\subsection{Einstein precession}
In this simplest case, all spin-related interactions are suppressed and only the classical 1PN orbital effects remain. Formally, the spin effects
are suppressed by setting $J_2=\tilde{G}=0$ and $\tilde{H}_\ast = H$ in the equations of motion \eqref{eq:averaged_momenta}-\eqref{eq:averaged_angles}.
We have to point out a formal difficulty in such a direct substitution, arising from the fact that when the spin $\tilde{G}$ is suppressed
the angles $h_{\ast}$ and $\tilde{h}$ become undefined and equations \eqref{eq:F1_H}-\eqref{eq:F1_Hts} have a pole in $\tilde{G}_{xy\ast}=0$\footnote{
This is analogous to the coordinates in the Delaunay set becoming undefined for zero inclination/eccentricity.
}. In order to avoid these problems,
we can write directly the equation of motion for $h=\tilde{h}+h_\ast$ in the general case as
\begin{align}
\frac{dh}{dt} & = \frac{d\tilde{h}}{dt} + \frac{d h_\ast}{dt} \notag \\
& = \epsilon\left[2\frac{{J_2}{\mathcal{G}}^{4}{m_2}^{3}}{{G}^{3}{L}^{3}}
-3\frac{{H}{J_2}{\mathcal{G}}^{4}{\tilde{H}_\ast}{m_2}^{3}}{{G}^{5}{L}^{3}}
-\frac{3}{2}\frac{{H}{\mathcal{G}}^{4}{m_2}^{4}}{{G}^{3}{L}^{3}}\right.\notag\\
&\quad+3\frac{{H}^{2}{J_2}{\mathcal{G}}^{4}{m_2}^{3}}{{G}^{5}{L}^{3}}
+\frac{3}{2}\frac{{\mathcal{G}}^{4}{\tilde{H}_\ast}{m_2}^{4}}{{G}^{3}{L}^{3}}+\left(\frac{3}{2}\frac{{H}^{2}{J_2}{\mathcal{G}}^{4}{\tilde{G}_{xy\ast}}{m_2}^{3}}{{G}^{5}{G_{xy}}{L}^{3}}\right.\notag\\
&\quad\left.\left.-\frac{3}{2}\frac{{G_{xy}}{J_2}{\mathcal{G}}^{4}{\tilde{G}_{xy\ast}}{m_2}^{3}}{{G}^{5}{L}^{3}}
-\frac{3}{2}\frac{{H}{\mathcal{G}}^{4}{\tilde{G}_{xy\ast}}{m_2}^{4}}{{G}^{3}{G_{xy}}{L}^{3}}\right)\cos{{h_\ast}}\right],
\end{align}
and verify that this equations is regular in absence of spins: the poles have disappeared and the the indeterminacy of $h_\ast$ is inconsequential as the
factor of $\cos{{h_\ast}}$ becomes zero in absence of spins.

The equations of motion of the mean orbital coordinates in absence of spins thus become
\begin{align}
\frac{dl}{dt} & =\frac{\mathcal{G}^{2}m_{2}^{2}}{L^{3}}+\epsilon\left(9\frac{{\mathcal{G}}^{4}{m_2}^{4}}{{G}{L}^{4}}-\frac{15}{2}\frac{{\mathcal{G}}^{4}{m_2}^{4}}{{L}^{5}}\right),\\
\frac{dg}{dt} & =3\epsilon\frac{{\mathcal{G}}^{4}{m_2}^{4}}{{G}^{2}{L}^{3}},\label{eq:ein_prec}\\
\frac{dh}{dt} & = 0,
\end{align}
whereas all mean momenta are constants of motion. We can recognize in eq. \eqref{eq:ein_prec} the classical relativistic
effect of the pericentre precession.
In Keplerian orbital elements (see conversion formul\ae{} \eqref{eq:Del_00}), eq. \eqref{eq:ein_prec} becomes:
\begin{equation}
\frac{dg}{dt}\equiv\frac{d\omega}{dt}=\frac{3m_{2}^{\frac{3}{2}}\mathcal{G}^{\frac{3}{2}}}{c^{2}a^{\frac{5}{2}}\left(1-e^{2}\right)},
\end{equation}
which, over an averaged orbit of the secondary body of period
\begin{equation}
T=2\pi\sqrt{\frac{a^{3}}{m_{2}\mathcal{G}}},
\end{equation}
amounts to
\begin{equation}
\Delta\omega=\frac{6\pi\mathcal{G}m_{2}}{c^{2}a\left(1-e^{2}\right)}.
\end{equation}
which is the classical formula for the relativistic pericentre precession \citep{ANDP:ANDP19163540702,straumann_general_1984}. The result
for the perturbation on the mean mean anomaly $l$ is in agreement with the results of
\citet{1988CeMec..43..193R} and \citet{heimberger_relativistic_1990}, also obtained
with a Lie series technique.
\subsection{Lense-Thirring effect}
The case in which only the central body is spinning corresponds to $J_{2}\neq0$, $\tilde{G}=0$
and $\tilde{H}_{\ast} = H$. As in the preceding case, all mean momenta are constants of motion,
and the equations
for the mean orbital coordinates become:
\begin{align}
\frac{dl}{dt} & =\frac{\mathcal{G}^{2}m_{2}^{2}}{L^{3}}+\epsilon\left(9\frac{{\mathcal{G}}^{4}{m_2}^{4}}{{G}{L}^{4}}-6\frac{{H}{J_2}{\mathcal{G}}^{4}{m_2}^{3}}{{G}^{3}{L}^{4}}-\frac{15}{2}\frac{{\mathcal{G}}^{4}{m_2}^{4}}{{L}^{5}}\right),\\
\frac{dg}{dt} & =\epsilon\left(-6\frac{{H}{J_2}{\mathcal{G}}^{4}{m_2}^{3}}{{G}^{4}{L}^{3}}+3\frac{{\mathcal{G}}^{4}{m_2}^{4}}{{G}^{2}{L}^{3}}\right),\\
\frac{dh}{dt} & = 2\epsilon\frac{{J_2}{\mathcal{G}}^{4}{m_2}^{3}}{{G}^{3}{L}^{3}}.
\end{align}
Here we can clearly recognize the effect of
the interaction of the spin of the central body with the orbit of the (non-rotating)
secondary body. This effect is sometimes referred to as Lense-Thirring
precession \citep{thirring_uber_1918}. The equations above show that the consequence
of endowing the central body with a spin is that of modifying Einstein's
pericentre precession via the additional term
\begin{equation}
-6\frac{{H}{J_2}{\mathcal{G}}^{4}{m_2}^{3}}{{G}^{4}{L}^{3}},
\end{equation}
and of introducing a precession of the mean line of nodes with angular
velocity
\begin{equation}
2\frac{{J_2}{\mathcal{G}}^{4}{m_2}^{3}}{c^2{G}^{3}{L}^{3}}.
\end{equation}
Both these two formul\ae{} are in agreement with the results of \citet{bogorodskii_relativistic_1959} and \citet{cugusi_relativistic_1978}.
\subsection{Geodetic effect}
\label{sec:geodetic_effect}
The case in which only the secondary body is spinning corresponds to $J_{2}=0$ and $\tilde{G}\neq0$.
This configuration is clearly more complicated than the previous two cases, as now $\mathcal{F}_1\neq 0$ and the mean momentum
$H$ is not a constant of motion any more. The equations of motion for $H$ and its conjugate mean coordinate $h_\ast$ read
\begin{align}
\frac{dH}{dt} & = \frac{3}{2}\epsilon\frac{{G_{xy}}{\mathcal{G}}^{4}{\tilde{G}_{xy\ast}}{m_2}^{4}}{{G}^{3}{L}^{3}}\sin{h_\ast},\label{eq:ge_00}\\
\frac{dh_\ast}{dt} & = \epsilon \left[ -3\frac{{H}{\mathcal{G}}^{4}{m_2}^{4}}{{G}^{3}{L}^{3}}
+\frac{3}{2}\frac{{\mathcal{G}}^{4}{\tilde{H}_\ast}{m_2}^{4}}{{G}^{3}{L}^{3}}
+\left(-\frac{3}{2}\frac{{G_{xy}}{H}{\mathcal{G}}^{4}{m_2}^{4}}{{G}^{3}{L}^{3}{\tilde{G}_{xy\ast}}}\right.\right.\notag\\
&\quad \left.\left.+\frac{3}{2}\frac{{G_{xy}}{\mathcal{G}}^{4}{\tilde{H}_\ast}{m_2}^{4}}{{G}^{3}{L}^{3}{\tilde{G}_{xy\ast}}}
-\frac{3}{2}\frac{{H}{\mathcal{G}}^{4}{\tilde{G}_{xy\ast}}{m_2}^{4}}{{G}^{3}{G_{xy}}{L}^{3}}\right)\cos{h_\ast}\right].\label{eq:ge_01}
\end{align}

A first observation is that, since $J_{2}=0$, in this case there is no preferred orientation for the reference system, and the form of the equations of motion
is rotationally invariant. As an immediate consequence, the fact that the mean momentum $\tilde{H}_\ast=H+\tilde{H}$
is constant regardless of the orientation of the reference system, implies that in this case the total mean angular momentum vector is a constant of motion
(whereas in the general case only its $z$ component is conserved). In turn, this observation,
together with the conservation of the norms of the mean orbital and rotational angular momenta $G$ and $\tilde{G}$, implies that the only possible
motion for the mean orbital and rotational angular momentum vectors is a simultaneous precession around the total mean angular momentum vector.

It is possible to verify this result and to quantify the precessional motion via a phase space analysis.
The dynamical system defined by equations \eqref{eq:ge_00}-\eqref{eq:ge_01}
has three equilibrium points. The first one,
\begin{equation}
\begin{aligned}
H^{\left(e\right)}&=\frac{G \tilde{H}_\ast}{G+\tilde{G}},\\
h_\ast^{\left(e\right)}&=2k\pi,
\end{aligned}
\end{equation}
with $k \in \mathbb{Z}$, corresponds to a geometrical configuration in which the mean orbital and rotational angular momentum
vectors are aligned and pointing in the same direction\footnote{
The geometrical interpretation of the equilibrium points can be verified with the help of eqs. \eqref{eq:Hts}-\eqref{eq:hs} and
Table \ref{tab:Ham_F_vars}.
}. Similarly, the second fixed point,
\begin{equation}
\begin{aligned}
H^{\left(e\right)}&=\frac{G \tilde{H}_\ast}{G-\tilde{G}},\\
h_\ast^{\left(e\right)}&=\pi+2k\pi,
\end{aligned}
\end{equation}
corresponds to a geometrical configuration in which the mean orbital and rotational angular momentum
vectors are aligned and pointing in opposite directions.
It is possible to check via direct substitution in eqs. \eqref{eq:averaged_angles} (with the help of
the partial derivatives in Appendix \ref{sec:full_form}) that, in both these two fixed points, the equation
of motion for $\tilde{h}$ collapses to
\begin{equation}
\frac{d\tilde{h}}{dt} = 0,
\end{equation}
thus implying that the angles $\tilde{h}$ and $h$ are constants. In other words, the mean spin and orbital angular momentum
vectors keep their mutually (anti) parallel configuration fixed in space.

The third and last fixed point,
\begin{equation}
\begin{aligned}
H^{\left(e\right)}&=\frac{\tilde{H}_\ast^2+G^2-\tilde{G}^2}{2 \tilde{H}_\ast},\\
h_\ast^{\left(e\right)}&=\pi+2 k \pi,
\end{aligned}
\end{equation}
corresponds to a geometrical configuration in which the projections of the mean spin and orbital angular momentum vectors on the $xy$ plane are
aligned, pointing in opposite directions and equal in magnitude. That is, the total mean angular momentum vector is lying on the $z$ axis.
In this fixed point, the equation of motion for both $h$ and $\tilde{h}$ reads
\begin{equation}
\frac{d\tilde{h}}{dt} = \frac{dh}{dt} = \frac{3  \mathcal{G}^4 m_2^4 \epsilon }{2 G^3 L^3}\tilde{H}_\ast.
\label{eq:ge_prec}
\end{equation}
In other words, both the orbital and rotational mean angular momenta of the secondary body precess around the total mean angular momentum vector
with an angular velocity given by eq. \eqref{eq:ge_prec}. Since, as remarked earlier, in absence of primary body spin the equations of motion
are rotationally invariant, it is always possible to orient the reference system
so that the total mean angular momentum vector is aligned to
the $z$ axis; the precessional motion described by this third fixed point will thus describe the evolution of mean spin and orbital angular momentum
in the general case.

The result in eq. \eqref{eq:ge_prec} agrees with the relatvistic effect known as \emph{geodetic} or \emph{de Sitter}
precession \citep{de_sitter_einsteins_1916}. In literature (see, e.g., \citet{Barker:1970vv,BARKER:1979wb} and
\citet{schiff1960experimental,schiff_motion_1960}) one finds the following formula
for the precession of the spin of a gyroscope orbiting a large (non-rotating)
spherical mass $m_{2}$ in a circular orbit:
\begin{equation}
\boldsymbol{\Omega}_{DS}^{\left(0\right)}=\frac{3\mathcal{G}\bar{\omega}m_{2}}{2c^{2}a}\boldsymbol{n},\label{eq:ds_barker}
\end{equation}
where $\bar{\omega}$ is the average orbital velocity
\begin{equation}
\bar{\omega}=\left(\frac{\mathcal{G}m_{2}}{a^{3}}\right)^{\frac{1}{2}},
\end{equation}
$a$ is the semi-major axis and $\boldsymbol{n}$ is a unit vector
in the direction of the orbital angular momentum. This formula agrees
with eq. \eqref{eq:ge_prec} in the small spin limit ($\tilde{H}_\ast \to H$) and for
circular orbits ($G \to L$).
\subsection{The general case}
We turn now our attention to the general case in which both bodies are spinning. First we will determine and characterise
the equilibrium points of the system, then we will study the exact solution of the equation of motion for $H$.
\subsubsection{Phase space analysis}
In the general case, Hamilton's equations for $H$ and $h_\ast$ are:
\begin{align}
\frac{dH}{dt} & =\epsilon\mathcal{F}_{1}\sin h_\ast,\label{eq:dHdt_00}\\
\frac{dh_{\ast}}{dt} &=\epsilon\left(\frac{\partial\mathcal{F}_{0}}{\partial H}+\frac{\partial\mathcal{F}_{1}}{\partial H}\cos h_\ast\right)\label{eq:dhsdt},
\end{align}
where $\mathcal{F}_{0}$ and $\mathcal{F}_{1}$ are given in eqs. \eqref{eq:F0}-\eqref{eq:F1}
and the partial derivatives are available in Appendix \ref{sec:full_form}.
It is clear that equilibrium points can be looked for in correspondence of either $\mathcal{F}_{1}=0$
or $\sin h_{\ast}=0$. We will consider these two cases separately.
\paragraph{Equilibria with $\mathcal{F}_{1}=0$}
The condition $\mathcal{F}_{1}=0$ implies the existence of the fixed point
\begin{align}
H^{\left(e\right)} &= \frac{G^{2}}{\mathcal{J}_{2}},\label{eq:F1_eq_00}\\
\cos h_{\ast}^{\left(e\right)} &=\frac{G^{2}\left(\frac{G^{2}}{\mathcal{J}_{2}}
+\mathcal{J}_{2}-\tilde{H}_{\ast}\right)}{\mathcal{J}_{2}\tilde{G}_{xy\ast}^{\left(e\right)}G_{xy}^{\left(e\right)}},
\label{eq:F1_eq_01}
\end{align}
where $G_{xy}^{\left(e\right)}$ and $\tilde{G}_{xy\ast}^{\left(e\right)}$
are $G_{xy}$ and $\tilde{G}_{xy\ast}$ evaluated in $H=H^{\left(e\right)}$, and we have defined
\begin{equation}
\mathcal{J}_{2}=\frac{J_{2}}{m_{2}}
\end{equation}
for notational convenience. The existence of this fixed point is subject to three constraints:
\begin{itemize}
\item since $G^{2}>0$ and $\mathcal{J}_{2}>0$ by definition, $H^{\left(e\right)}$
must also be positive;
\item since $H\leq G$ by definition (as $H$ is the $z$ component of the
mean specific orbital angular momentum, whose magnitude is $G$), the equilibrium
can exist only if $G\leq\mathcal{J}_{2}$;
\item finally, since $\cos h_{\ast}^{\left(e\right)}$ must assume values
in the $\left[-1,1\right]$ interval, eq. \eqref{eq:F1_eq_01} implies constraints on the values of the
mean momenta.
\end{itemize}
In order to characterise this equilibrium point, we can compute the eigenvalues $\lambda$ of the Jacobian in the equilibrium point, and obtain
\begin{equation}
\lambda=\pm\frac{3}{2}\epsilon\frac{m_{2}^{4}\mathcal{G}^{4}}{L^{3}G^{5}}
\tilde{G}_{xy\ast}^{\left(e\right)}G_{xy}^{\left(e\right)}\mathcal{J}_{2}\sin h_{\ast}^{\left(e\right)}.
\end{equation}
The eigenvalues are real and of opposite signs, and
the fixed point is thus a saddle\footnote{
This is of course an expected result,
since equilibrium points in one degree-of-freedom Hamiltonian systems must either be
centers,
saddles or have a double zero eigenvalue \citep{arnold_mathematical_1989}.}.

Note that, due to the structure of eq. \eqref{eq:dHdt_00}, the initial condition
$H=H^{\left(e\right)}$ results in $dH/dt=0$ regardless
of the initial value of $h_{\ast}$. As a consequence, the line $H=H^{\left(e\right)}$
is an invariant submanifold of the phase space. We can characterise explicitly the time behaviour of  $h_{\ast}$ on this submanifold.
For an arbitrary initial value of $h_{\ast}$, we can solve eq. \eqref{eq:dhsdt} by quadrature
and obtain
\begin{equation}
h_{\ast}\left(t\right)=2\tan^{-1}\left[\frac{\sqrt{B^{2}-A^{2}}}{A+B}\tanh\left(\frac{\left(C-t\right)\sqrt{B^{2}-A^{2}}}{2}\right)\right],\label{eq:hs_t}
\end{equation}
where 
\begin{align}
A & =\frac{3}{2}\epsilon\frac{m_{2}^{4}\mathcal{G}^{4}}{L^{3}G^{3}}\left(\frac{G^{2}}{\mathcal{J}_{2}}+\mathcal{J}_{2}-\tilde{H}_{\ast}\right),\label{eq:A_def}\\
B & =\frac{3}{2}\epsilon\frac{m_{2}^{4}\mathcal{G}^{4}}{L^{3}G^{5}}\tilde{G}_{xy\ast}^{\left(e\right)}G_{xy}^{\left(e\right)}\mathcal{J}_{2}.\label{eq:B_def}
\end{align}
Note that $B$ is non-negative by definition. 
We can distinguish different behaviours on the line $H=H^{\left(e\right)}$ depending on the values of $A$ and $B$.

In the first case, $B>\left|A\right|$ and the radicands in eq. \eqref{eq:hs_t}
are positive, so that
\begin{equation}
\lim_{t\to\infty}h_{\ast}\left(t\right)=-2\tan^{-1}\left(\frac{\sqrt{B^{2}-A^{2}}}{A+B}\right).
\end{equation}
Noting from eqs. \eqref{eq:F1_eq_01}, \eqref{eq:A_def} and \eqref{eq:B_def} that
\begin{equation}
\cos h_{\ast}^{\left(e\right)}=\frac{A}{B},
\end{equation}
we have
\begin{equation}
\frac{\sqrt{B^{2}-A^{2}}}{A+B}=\left|\tan\frac{h_{\ast}^{\left(e\right)}}{2}\right|,
\end{equation}
and hence
\begin{equation}
\cos\left(\lim_{t\to\infty}h_{\ast}\left(t\right)\right)=\cos h_{\ast}^{\left(e\right)}.
\end{equation}
That is, when $B>\left|A\right|$ $h_{\ast}\left(t\right)$ tends
asymptotically to the equilibrium and we can speak of an attractor on the invariant submanifold $H=H^{\left(e\right)}$.

In the second case $B<\left|A\right|$ and the square roots in eq. \eqref{eq:hs_t}
are purely imaginary. Thanks to elementary properties of hyperbolic
functions, we can then write
\begin{align}
h_{\ast}\left(t\right) & =2\tan^{-1}\left[\frac{\sqrt{A^{2}-B^{2}}}{A+B}\tan\left(\frac{\left(t-C\right)\sqrt{A^{2}-B^{2}}}{2}\right)\right].
\end{align}
That is, the evolution of $h_{\ast}$ in time is linear with the superimposition
of a periodic modulation.

In the last case, $\left|A\right|=B$ and eq. \eqref{eq:hs_t} simplifies
into two different formul\ae{} depending on the sign of $A$.

If $A>0$, the time evolution of $h_{\ast}$ is given by
\begin{equation}
h_{\ast}\left(t\right)=2\tan^{-1}\left(\frac{1}{C-Bt}\right),
\end{equation}
and $h_{\ast}$ tends asymptotically to $0$.

If $A<0$, the time evolution of $h_{\ast}$ is given by
\begin{equation}
h_{\ast}\left(t\right)=2\tan^{-1}\left(-Bt-C\right),
\end{equation}
and $h_{\ast}$ tends asymptotically to $\pi$.

The analysis above clearly shows that it is possible to prepare the system
in such a way to induce an aperiodic dynamical evolution. 
However, the substitution of values for $G$, $\tilde{G}$,  $\tilde{H}_{\ast}$ and $\mathcal{J}_{2}$
typical of planetary systems (even in the case of pulsar planetary systems)
shows that the orbits in correspondence of the equilibrium point have a very small semi-major
axis which either puts the orbit inside the main body or in a zone in which the approximation
of weak field starts to fail.
In this respect then the equilibrium condition
is interesting, but for a meaningful physical interpretation an analysis at higher PN orders
and/or within the framework of the full two-body problem is required.
\paragraph{Equilibria with $\sin h_\ast = 0$}
The case $\sin h_{\ast}=0$, similarly to the secondary body spin
configuration, implies $h_{\ast}^{\left(e\right)}=k\pi$ (with $k\in\mathbb{Z}$),
and leads to the following equation of motion for $h_{\ast}$:
\begin{equation}
\frac{dh_{\ast}}{dt}\left(H,k\pi\right)  =\frac{3}{2}\epsilon\frac{m_{2}^{4}\mathcal{G}^{4}}{L^{3}G^{3}}\left[\mathcal{R}_{1}\left(H\right)\pm\mathcal{R}_{2}\left(H\right)\right],
\end{equation}
where 
\begin{align}
\mathcal{R}_{1}\left(H\right) & =-2\frac{H\tilde{H}_{\ast}\mathcal{J}_{2}}{G^{2}}-2H+3\frac{H^{2}\mathcal{J}_{2}}{G^{2}}+\tilde{H}_{\ast}+\mathcal{J}_{2},\\
\mathcal{R}_{2}\left(H\right) & =\frac{H^{2}G_{xy}\mathcal{J}_{2}}{\tilde{G}_{xy\ast}G^{2}}
-\frac{\tilde{G}_{xy\ast}G_{xy}\mathcal{J}_{2}}{G^{2}}
-\frac{H\tilde{G}_{xy\ast}}{G_{xy}}+\frac{H^{2}\tilde{G}_{xy\ast}\mathcal{J}_{2}}{G_{xy}G^{2}}\notag\\
&\quad -\frac{HG_{xy}}{\tilde{G}_{xy\ast}}+\frac{\tilde{H}_{\ast}G_{xy}}{\tilde{G}_{xy\ast}}
-\frac{H\tilde{H}_{\ast}G_{xy}\mathcal{J}_{2}}{\tilde{G}_{xy\ast}G^{2}},
\end{align}
and the $\pm$ depends on $k$. The equilibrium condition becomes then
\begin{equation}
\mathcal{R}_{1}\left(H\right)=\pm\mathcal{R}_{2}\left(H\right).\label{eq:gen_eq_cond}
\end{equation}
Recalling that $G_{xy}$ and $\tilde{G}_{xy\ast}$ are functions of
$H$, we can multiply both sides of eq. \eqref{eq:gen_eq_cond} by
$\tilde{G}_{xy\ast}G_{xy}$, square them and, after straightforward
but tedious calculations, obtain the equilibrium condition
\begin{equation}\label{eq:pol_6}
f_{6}\left(H\right)=0,
\end{equation}
where $f_{6}\left(H\right)$ is a polynomial of degree six in $H$
whose coefficients are functions of the constants of motion. The equilibrium
points for $H$ will be zeroes of this polynomial, but not all zeroes
constitute valid equilibrium points: non-real zeroes, and zeroes that
do not respect the condition $\left|H\right|\leq G$ have to be discarded.

A noteworthy particular equilibrium point is $h_{\ast}^{\left(e\right)}=H^{\left(e\right)}=0$,
with $\tilde{H}_{\ast}=0$ and $G=\tilde{G}$. This geometrical configuration
corresponds to a polar orbit in which the mean spin vector of the secondary
body is identical (in both magnitude and orientation) to the mean orbital
angular momentum vector. From eqs. \eqref{eq:averaged_momenta} it
follows immediately that, while $h_{\ast}$ remains constantly zero,
$\tilde{h}$ and $h$ rotate with constant angular velocity
\begin{equation}
\frac{1}{2}\epsilon\frac{m_{2}^{3}\mathcal{G}^{4}J_{2}}{L^{3}G^{3}}.
\end{equation}

\begin{figure}%[htbp]
\begin{center}
\includegraphics[scale=0.25]{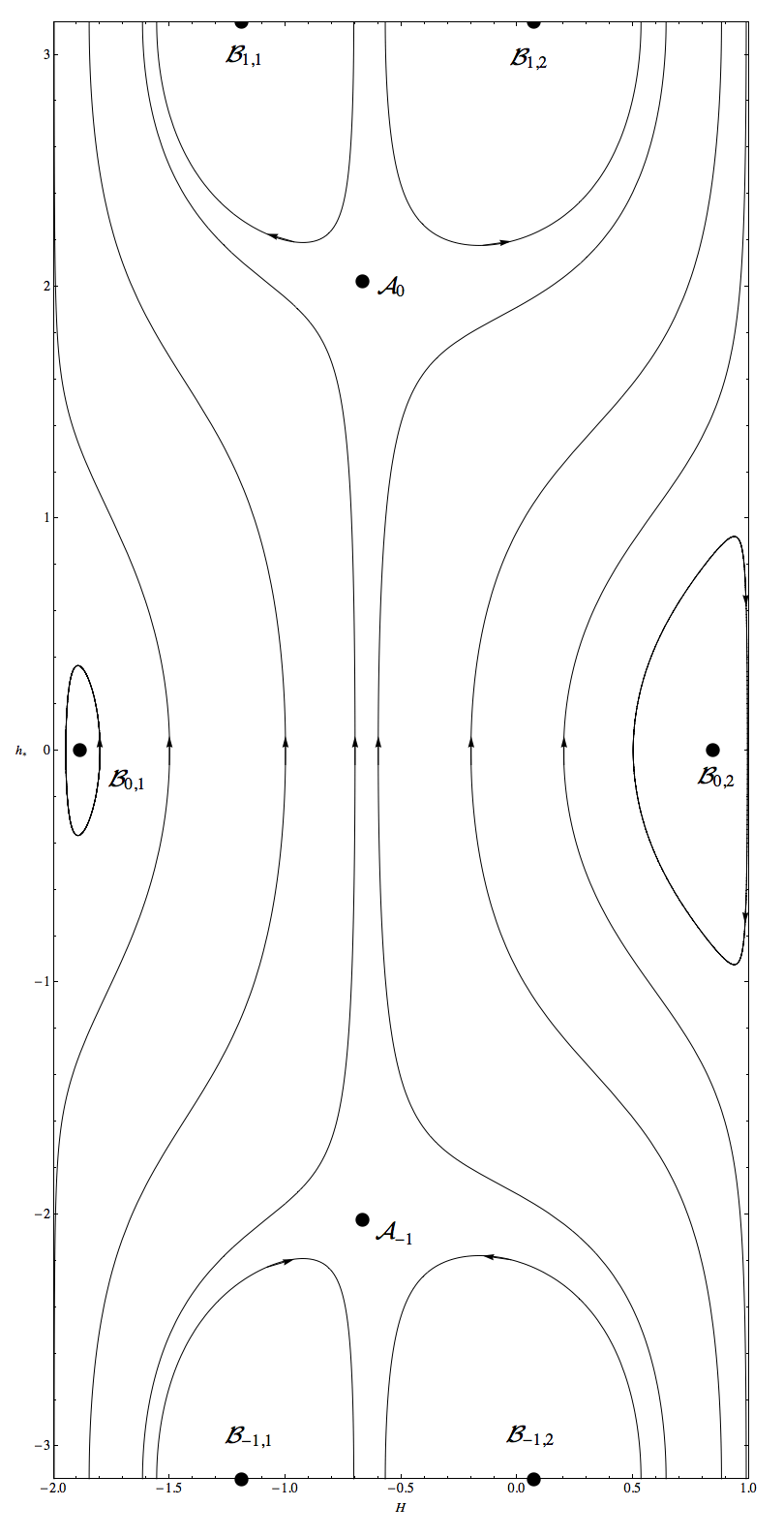}
\caption{Sample of the phase space of the system \eqref{eq:dHdt_00}-\eqref{eq:dhsdt}.
The line connecting the points $\mathcal{A}_0$ and $\mathcal{A}_1$ is an invariant submanifold. On this submanifold, $\mathcal{A}_0$ is an attractor.}
\label{fig:ph_sp_gen}
\end{center}
\end{figure}

The results above indicate that the phase space has at most six fixed points, but our inability to solve 
\eqref{eq:pol_6} in the general case prevents us from giving a complete general analysis.
Thus, here we limit ourselves to show,
in Figure \ref{fig:ph_sp_gen}, a sample of the phase space for a specific choice of the parameters,
in order to give an idea of the kind of structures that one is likely to encounter in this analysis.
\subsubsection{Exact solution for $H\left(t\right)$}
We now derive an analytical solution for the time variation of $H$.
The availability of a closed-form solution for $H\left(t\right)$
allows to obtain immediately the time evolution of $\cos h_{\ast}$
via inversion of the Hamiltonian \eqref{eq:Ham_av_00}\footnote{This, of course, 
is not possible if singular equilibrium points  are present or if we are dealing with the 
indeterminate forms arising when the nodal angles are undefined.}.
With $H\left(t\right)$ and $\cos h_{\ast}\left(t\right)$ it is then
possible in principle to integrate the equations of motion for the
remaining coordinates. Additionally, the analytical expression will
allow us to calculate exactly the period of the oscillatory motion of $H\left(t\right)$
and to make quantitative predictions about the behaviour of real physical systems in \S\ref{sec:physical_systems}.

Recalling the form of the Hamiltonian \eqref{eq:Ham_av_00}  and combining it with the
equation for $H$,
\begin{equation}
\frac{dH}{dt}=\epsilon\mathcal{F}_{1}\sin h_\ast,\label{eq:H_integral}
\end{equation}
we have
\begin{equation}
\frac{dH}{dt}=\pm\sqrt{\epsilon^{2}\mathcal{F}_{1}^{2}-\left(\mathcal{H}^{\prime}-\mathcal{H}_{N}-\epsilon\mathcal{F}_{0}\right)^{2}},\label{eq:dHdt}
\end{equation}
where $\mathcal{H}^\prime$ is the Hamiltonian constant
(whose value is computed by substituting the initial values of the canonical
variables in eq. \eqref{eq:Ham_av_00}),
and with the understanding that the plus sign is now to be taken when
$\epsilon\mathcal{F}_{1}$ and $\sin h_{\ast}$ have the same sign.
At this point it can be easily proven  by direct calculations that the terms
in $H^6$ and $H^5$ cancel out so that the radicand in eq. \eqref{eq:dHdt} is a quartic polynomial in $H$.
Following \citet{WhWa27}, we can then elect
\begin{align}
f_{4}\left(H\right) & =\epsilon^{2}\mathcal{F}_{1}^{2}-\left(\mathcal{H}^{\prime}-\mathcal{H}_{N}-\epsilon\mathcal{F}_{0}\right)^{2}\nonumber \\
 & =a_{4}+4a_{3}H+6a_{2}H^{2}+4a_{1}H^{3}+a_{0}H^{4},\label{eq:f4H_def}
\end{align}
and rewrite eq. \eqref{eq:dHdt} as 
\begin{equation}
\int_{H_{0}}^{H}\pm\frac{dx}{\sqrt{f_{4}(x)}}=\int_{t_{0}}^{t}d\tau,\label{eq:ell_inv_H}
\end{equation}
where $H_{0}$ is the initial value of $H$ and $t_{0}$ the initial
time. The coefficients of $f_{4}\left(H\right)$ are reproduced in full form
in Appendix \ref{sec:cf_f4H}.
The left-hand side of this equation is, apart from the sign
ambiguity, an elliptic integral in standard form. A formula by Weierstrass
(see \citet{WhWa27}, \S 20.6) allows to write the upper
limit of the integral on the left-hand side as a function of the right-hand
side. Specifically, if
\begin{equation}
z=\int_{a}^{x}\left\{ f_{4}\left(t\right)\right\} ^{-\frac{1}{2}}dt,\label{eq:weier_int}
\end{equation}
where $a$ is any constant, then
\begin{align}
x\left(z\right) & = a + \frac{1}{2\left[\wp\left(z\right)-\frac{1}{24}f_{4}^{\prime\prime}\left(a\right)\right]^{2}
-\frac{1}{48}f_{4}\left(a\right)f_{4}^{iv}\left(a\right)}\notag\\
&\quad\cdot\left\{\frac{1}{2}f_{4}^{\prime}\left(a\right)\left[\wp\left(z\right)
-\frac{1}{24}f_{4}^{\prime\prime}\left(a\right)\right]+\frac{1}{24}f_{4}\left(a\right)f_{4}^{\prime\prime\prime}\left(a\right)
\right.\notag\\
&\quad\left.+\sqrt{f_{4}\left(a\right)}\wp^{\prime}\left(z\right)
\vphantom{\frac{1}{2}f_{4}^{\prime}\left(a\right)\left[\wp\left(z\right)
-\frac{1}{24}f_{4}^{\prime\prime}\left(a\right)\right]}
\right\},
\label{eq:weier_inv}
\end{align}
where $\wp\left(z\right)\equiv\wp\left(z;g_{2},g_{3}\right)$ is a
Weierstrass elliptic function defined in terms of the invariants
\begin{align}
g_{2} & =a_{0}a_{4}-4a_{1}a_{3}+3a_{2}^{2},\\
g_{3} & =a_{0}a_{2}a_{4}+2a_{1}a_{2}a_{3}-a_{2}^{3}-a_{0}a_{3}^{2}-a_{1}^{2}a_{4},
\end{align}
and the derivatives of $f_{4}$ are intended with respect to the polynomial variable.
After the removal of the sign ambiguity in eq. \eqref{eq:ell_inv_H}, detailed in Appendix \ref{sec:sign_removal},
the time evolution of $H$ is thus given by
\begin{align}
H\left(t\right)& = H_{0}+\frac{1}
{2\left[\wp\left(t\right)-\frac{1}{24}f_{4}^{\prime\prime}\left(H_{0}\right)\right]^{2}-\frac{1}{48}f_{4}\left(H_{0}\right)f_{4}^{iv}\left(H_{0}\right)}\notag\\
&\quad\cdot\left\{\frac{1}{2}f_{4}^{\prime}\left(H_{0}\right)\left[\wp\left(t\right)
-\frac{1}{24}f_{4}^{\prime\prime}\left(H_{0}\right)\right]
\right.\notag\\
&\quad\left.+\frac{1}{24}f_{4}\left(H_{0}\right)f_{4}^{\prime\prime\prime}\left(H_{0}\right)
\pm\sqrt{f_{4}\left(H_{0}\right)}\wp^{\prime}\left(t\right)\vphantom{
\frac{1}{2}f_{4}^{\prime}\left(H_{0}\right)\left[\wp\left(t\right)
-\frac{1}{24}f_{4}^{\prime\prime}\left(H_{0}\right)\right]
}\right\},
\label{eq:H_of_t}
\end{align}
where the $\pm$ sign is chosen based on the initial signs of $\sin h_{\ast}$
and $\mathcal{F}_{1}$, and we set the initial time $t_{0}=0$ for convenience.

It is now useful
to recall some basic properties of $\wp\left(t\right)$ \citep{Abramowitz}. 
First of all, since in our case the invariants $g_{2}$
and $g_{3}$ are defined to be purely real and $t$ is also limited
to real values, $\wp\left(t\right)$ and $H\left(t\right)$ become
real-valued periodic functions. The period of $\wp\left(t\right)$,
$\wp^{\prime}\left(t\right)$ and $H\left(t\right)$ is related to
the invariants $g_{2}$ and $g_{3}$ via formul\ae{} involving elliptic
integrals and the roots $e_{1}$, $e_{2}$ and $e_{3}$ of the cubic
equation
\begin{equation}
4t^{3}-g_{2}t-g_{3}=0.\label{eq:cubic_eq}
\end{equation}
The sign of the modular discriminant
\begin{equation}
\Delta=g_{2}^{3}-27g_{3}^{2},\label{eq:mod_discri}
\end{equation}
determines the nature of the roots $e_{1}$, $e_{2}$ and $e_{3}$
of the cubic equation \eqref{eq:cubic_eq} and thus also the value
of the period.

$\wp\left(t\right)$ can degenerate into a non-periodic function when
$\Delta=0$. In particular, if $g_{2}=g_{3}=0$, $\wp\left(t\right)$
becomes simply
\begin{equation}
\wp\left(t\right)=\frac{1}{t^{2}}.
\end{equation}
If $g_{2}>0$ and $g_{3}<0$, the case $\Delta=0$ degenerates instead
to
\begin{equation}
\wp\left(t\right)=e_{1}+\frac{3e_{1}}{\left\{ \sinh\left[\left(3e_{1}\right)^{\frac{1}{2}}t\right]\right\} ^{2}},\label{eq:wp_simpl_nonperiodic}
\end{equation}
where $e_{1}=e_{2}>0$ and $e_{3}=-2e_{1}$. For $g_{2}>0$ and $g_{3}>0$,
the case $\Delta=0$ simplifies to
\begin{equation}
\wp\left(t\right)=-\frac{e_{1}}{2}+\frac{\frac{3}{2}e_{1}}{\left\{ \sin\left[\left(\frac{3}{2}e_{1}\right)^{\frac{1}{2}}t\right]\right\} ^{2}},\label{eq:wp_simpl_periodic}
\end{equation}
where $e_{1}>0$ and $e_{2}=e_{3}<0$. When $\wp\left(t\right)$ is
non-periodic, for $t\to\infty$ $H\left(t\right)$ tends to a finite
value.

In addition to the degenerate cases of $\wp\left(t\right)$, $H\left(t\right)$
can degenerate also via particular values of $H_{0}$. For instance,
if $f_{4}\left(H_{0}\right)=f_{4}^{\prime}\left(H_{0}\right)=0$,
it is then clear from eq. \eqref{eq:H_of_t} that $H\left(t\right)$
degenerates to
\begin{equation}
H\left(t\right)=H_{0},
\end{equation}
which corresponds to an equilibrium point for $H$. It is straightforward,
although tedious if done by hand, to check by substitution that the
equilibrium condition found in eq. \eqref{eq:F1_eq_00}
, $H_{0}=G^{2}/\mathcal{J}_{2}$, leads to the condition $f_{4}\left(H_{0}\right)=f_{4}^{\prime}\left(H_{0}\right)=0$.
Another equilibrium condition that can be checked via substitution
is the geometric configuration in which both the mean orbital angular momentum
and the secondary body's mean spin are initially aligned to the spin of
the primary, i.e., $H_{0}=G$ and $\tilde{H}_{\ast}=G+\tilde{G}$.
This approach allows us to sidestep difficulties in verifying this
equilibrium condition (which results in indeterminate forms due to
$G_{xy}=\tilde{G}_{xy\ast}=0$ and $h_{\ast}$ being undefined) within
the dynamical systems framework.

As a special case of the general formula \eqref{eq:H_of_t}, we can
analyse the behaviour of $H\left(t\right)$ when $J_{2}=0$ (i.e.,
the particular case of geodetic precession examined in \S\ref{sec:geodetic_effect}).
It is easy to check by substitution (see formul\ae{} in Appendix \ref{sec:cf_f4H})
that in this case the coefficients
$a_{0}$ and $a_{1}$ of the quartic polynomial $f_{4}\left(H\right)$
are both zero, and thus
\begin{align}
g_{2} & =3a_{2}^{2},\\
g_{3} & =-a_{2}^{3},
\end{align}
with
\begin{align}
a_{2} & =-\frac{1}{2}\epsilon\frac{m_{2}^{4}\mathcal{G}^{4}\mathcal{H}^{\prime}}{G^{3}L^{3}}
+\frac{15}{16}\epsilon^{2}\frac{m_{2}^{8}\mathcal{G}^{8}}{G^{3}L^{7}}
+\frac{1}{4}\epsilon\frac{\tilde{G}^{2}m_{2}^{4}\mathcal{G}^{4}\mathcal{I}_{1}}{G^{3}L^{3}}\notag\\
&\quad-\frac{15}{8}\epsilon^{2}\frac{m_{2}^{8}\mathcal{G}^{8}}{G^{4}L^{6}}
-\frac{3}{8}\epsilon^{2}\frac{\tilde{G}^{2}m_{2}^{8}\mathcal{G}^{8}}{G^{6}L^{6}}
-\frac{1}{4}\epsilon\frac{m_{2}^{6}\mathcal{G}^{6}}{G^{3}L^{5}},
\end{align}
and the modular discriminant $\Delta$ is also null. As explained
earlier and depending on the sign of $g_{3}$, in this configuration
$\wp\left(t\right)$ is either a simplified periodic function of the
form \eqref{eq:wp_simpl_periodic} or a non-periodic function of the
form \eqref{eq:wp_simpl_nonperiodic}. However, it is easy to check by
substitution of $\mathcal{H}^{\prime}$ that
\begin{align}
a_{2} & = -\frac{3}{8}\epsilon^2\frac{m_2^8\mathcal{G}^8}{L^6G^6}\left[G^2-2H_0^2+2H_0\tilde{H}_\ast+\tilde{G}^2+\right.\notag\\
&\quad\left.2G_{xy,0}\tilde{G}_{xy,0}\cos h_{\ast,0}\right],
\end{align}
where the zero subscript represents initial values. It is straightforward to verify
how the quantity in the square brackets is the square of the magnitude $M$ of the total mean angular momentum vector, and therefore it cannot be a negative quantity.
As a first consequence, $g_3$ is always positive and the behaviour of $H\left(t\right)$ is restricted to be purely periodic. Secondly, according
to eq. \eqref{eq:wp_simpl_periodic} and keeping in mind that $e_1=-a_2$, the angular velocity of the periodic motion of $H\left(t\right)$ is
\begin{equation}
2\sqrt{\frac{3}{2}e_1} = 2\sqrt{\frac{9}{16}\epsilon^2\frac{m_2^8\mathcal{G}^8}{L^6G^6}M^2} = \frac{3}{2}\epsilon\frac{m_2^4\mathcal{G}^4}{L^3G^3}M
\label{eq:ge_from_pe}
\end{equation}
(the factor 2 arising from the fact that in \eqref{eq:wp_simpl_periodic} the sine is squared),
which is in agreement with the precessional angular velocity \eqref{eq:ge_prec} calculated for the geodetic effect\footnote{
In formula \eqref{eq:ge_prec}, the $z$ component of the total mean angular momentum $\tilde{H}_\ast$ coincides with $M$
in formula \eqref{eq:ge_from_pe}, as in eq. \eqref{eq:ge_prec} the
total mean angular momentum is aligned to the $z$ axis.
}.
\section{Application to physical systems}
\label{sec:physical_systems}
We are now ready to apply the machinery we have developed so far to some real physical systems.
We will consider three examples of interest: a Mercury-like planet,
an exoplanet in close orbit around a pulsar and the Earth-orbiting Gravity Probe B experiment.
It is clear that, because of our initial assumptions, we will consider highly idealised systems;
we have to stress again that our goal here is not to produce accurate predictions of the dynamical
evolution of realistic physical models, but rather to highlight the role that relativistic effects
play in the long term.

We will find  that in all cases the spin-orbit and spin-spin couplings induce periodic oscillations of the mean orbital plane and,
at the same time, of the mean spin vector. Such effects are typically small for the orbital plane, but,
because of the conservation of the $z$ component of the total angular momentum $\tilde{H}_\ast$,
they correspond to non-negligible oscillations in the orientation of the mean spin.

We have to point out
that the perturbative treatment outlined in the previous sections produces results in terms of the components
of the mean angular momentum vectors with respect to a fixed (non-rotating)
centre-of-mass reference system (rather
than in terms of obliquity and relative orientations).
Therefore, in the geometrical interpretations of our results,
we will always be referring to \emph{absolute} (as opposed to \emph{relative})
orientation angles.

Given the small magnitude of the quantities involved in the computations, in order to avoid numerical issues
we performed all calculations using the {\it mpmath} multiprecision library \citep{mpmath}.
\subsection{Mercury-like planet}
In the simplified case in which the Sun is a rigid homogeneous sphere
rotating around a fixed axis, and a Mercury-like
planet is the only body orbiting it, a set of possible initial values
for the parameters of the system are displayed in Table \ref{tab:merc_params}.
The parameters correspond to a Mercury-sized object orbiting at $0.47\ \mathrm{AU}$
from the Sun on an orbit with eccentricity $0.2$ and inclination
$7^{\circ}$. The plane of the ecliptic is identified as being perpendicular
to the spin axis of the Sun. The planet rotates with a period of $1405\:\mathrm{h}$,
and the absolute inclination of its spin vector is close to the value of the orbital inclination.
Its radius $r_{1}$ is $2439\:\mathrm{km}$.
\begin{table}
\begin{center}
\begin{tabular}{cc}
\toprule 
Parameter & Value (SI units)\tabularnewline
\midrule
$L_{0}$ & $2.77\times10^{15}$\tabularnewline
$G_{0}$ & $2.71\times10^{15}$\tabularnewline
$H_{0}$ & $2.69\times10^{15}$\tabularnewline
$\tilde{G}_{0}$ & $2.95\times10^{6}$\tabularnewline
$\tilde{H}_{0}$ & $2.93\times10^{6}$\tabularnewline
$J_{2}$ & $1.12\times10^{42}$\tabularnewline
$r_{1}$ & $6.37\times10^{6}$\tabularnewline
\bottomrule
\end{tabular}
\end{center}
\caption{Initial values (in SI units) for the parameters of a simplified Sun-Mercury
two-body system.\label{tab:merc_params}}
\end{table}
\begin{figure*}
\begin{center}
\includegraphics[scale=0.60]{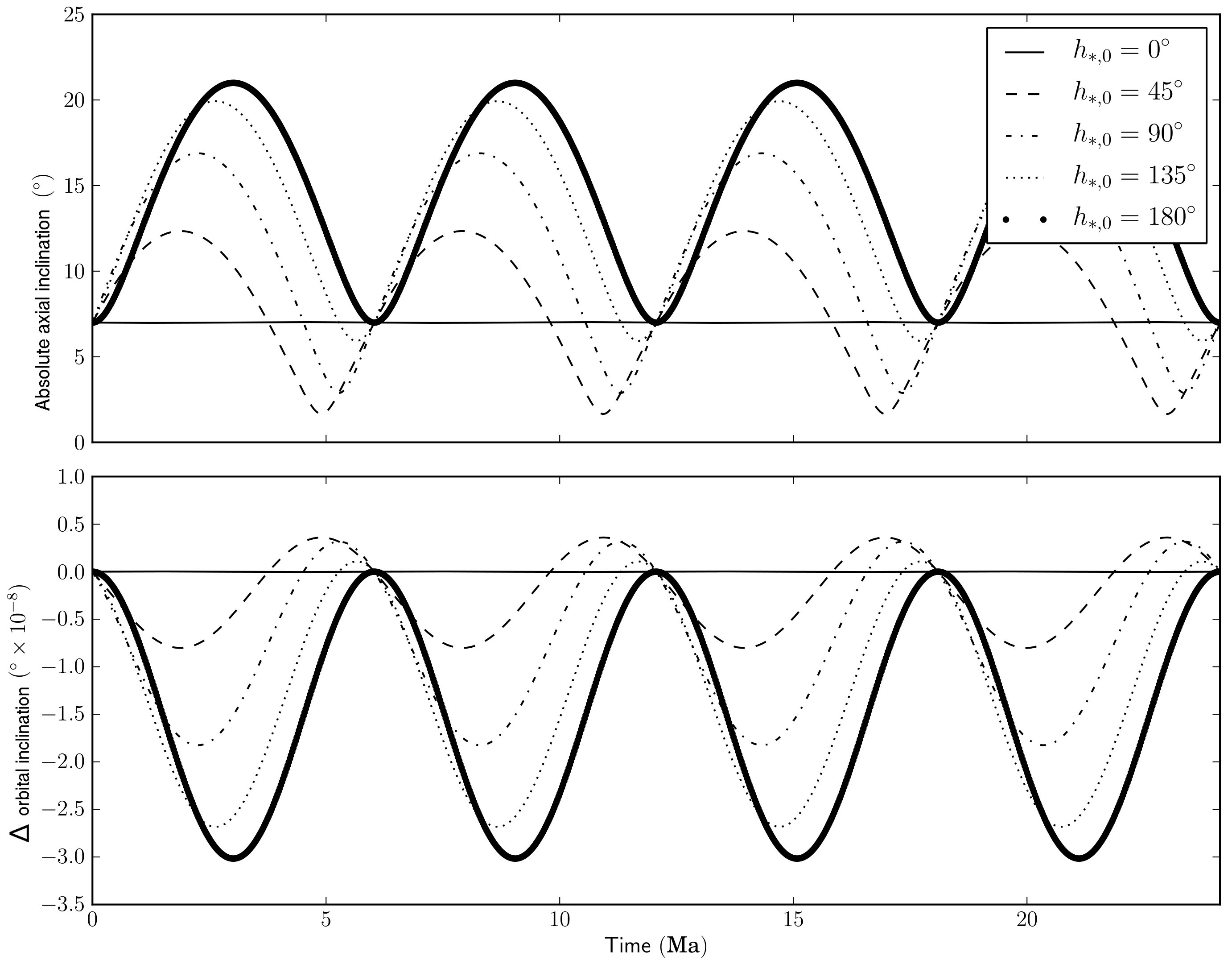}
\end{center}
\caption{Time evolution of the absolute axial (top) and orbital (bottom) inclinations
of a Mercury-like object orbiting the Sun. In the bottom panel the
quantity on the $y$ axis is the difference from the initial value
of orbital inclination. The different curves correspond to different
initial values for $h_{\ast}$. Time is measured in millions of years.\label{fig:merc_plot}}
\end{figure*}

Figure \ref{fig:merc_plot} shows the time evolution of the absolute axial
and orbital inclinations of the planet for different initial
values of the angle $h_{\ast}$. The period of the evolution, calculated
from the solution for $H\left( t \right)$ in terms of elliptic functions,
is $6.03\:\mathrm{Ma}$.
When $h_{\ast,0}=0$ the system is almost in equilibrium, as the initial
axial and orbital inclinations are almost equal and almost
parallel to the Sun's spin vector. The amplitude of the periodic oscillation
increases together with $h_{\ast,0}$. The oscillation is much wider
in axial than in orbital inclination; this is a consequence of
the fact that for this system $G\gg\tilde{G}$: the conservation of
$\tilde{H}_{\ast}=H+\tilde{H}$, $G$ and $\tilde{G}$ imposes that
a small change in the $z$ component of the mean orbital angular momentum,
$H$, is proportionally a much larger change in the $z$ component
of the mean spin vector, $\tilde{H}$.

The main effect that can be observed in Figure \ref{fig:merc_plot} is the geodetic precession.
Indeed, in this dynamical system the slow rotations of both the planet and the Sun minimise the spin-spin interactions,
and effectively relegate the spin-orbit effects to a precession of the planet's mean spin axis around the mean orbital
angular momentum vector. This can be verified by noting that the precessional rate given by eq. \eqref{eq:ge_prec} yields
essentially the same period of $6.03\:\mathrm{Ma}$ as the general formula in terms of elliptic functions.

The correlation between the oscillation amplitude and $h_{\ast,0}$ has a simple geometrical interpretation:
when $h_{\ast,0} = 0$, the mean spin and orbital angular momentum vectors share the same inclination ($7^{\circ}$) \emph{and}
nodal angle, and therefore they are parallel (i.e., their relative angular separation is zero) and no precession motion
takes place; as the difference in initial nodal angles (i.e., $h_{\ast,0}$) increases, the relative angular separation between the two
vectors increases too and the mean spin precesses around the mean orbital angular momentum vector. When $h_{\ast,0} = \pi$, the
initial angular separation reaches the maximum possible value, and the projection of the precession motion on the $z$ axis
(which is what is visualised in Figure \ref{fig:merc_plot}) reaches its maximum oscillatory amplitude too.
\subsection{Pulsar planet}
In this second case, the central body is a millisecond pulsar with
a Jupiter-like planet in close orbit. The parameters of the system
are displayed in Table \ref{tab:psr_params}, and they are similar
to the estimated parameters of the PSR J1719-1438 system \citep{Bailes23092011}:
the mass of the star is $1.4\:\mathrm{M}_{\odot}$, its spin period
is $5.8\ \mathrm{ms}$ and its diameter is $20\ \mathrm{km}$, while
the planet has a mass roughly equal to Jupiter ($1.02\ \mathrm{M}_{\jupiter}$)
and a radius of $0.4\: r_{\jupiter}$, orbiting on a moderately-inclined
($i=20{}^{\circ}$) near-circular ($e=0.06$) orbit with semi-major
axis of $600000\:\mathrm{km}$. Lacking estimates on the rotational
state of the planet, we hypothesize a spin period similar to Jupiter
($10\ \mathrm{h}$) and an absolute inclination of the spin vector of $10^{\circ}$.
As in the preceding case, the plane of the ecliptic is identified as being perpendicular
to the spin axis of the star.
\begin{table}
\begin{center}
\begin{tabular}{cc}
\toprule 
Parameter & Value (SI units)\tabularnewline
\midrule
$L_{0}$ & $3.33\times10^{14}$\tabularnewline
$G_{0}$ & $3.32\times10^{14}$\tabularnewline
$H_{0}$ & $3.12\times10^{14}$\tabularnewline
$\tilde{G}_{0}$ & $5.32\times10^{10}$\tabularnewline
$\tilde{H}_{0}$ & $5.23\times10^{10}$\tabularnewline
$J_{2}$ & $4.83\times10^{41}$\tabularnewline
$r_{1}$ & $2.76\times10^{7}$\tabularnewline
\bottomrule
\end{tabular}
\end{center}
\caption{Initial values (in SI units) for the parameters of a pulsar-Jovian
planet two-body system.\label{tab:psr_params}}
\end{table}
\begin{figure*}
\begin{center}
\includegraphics[scale=0.54]{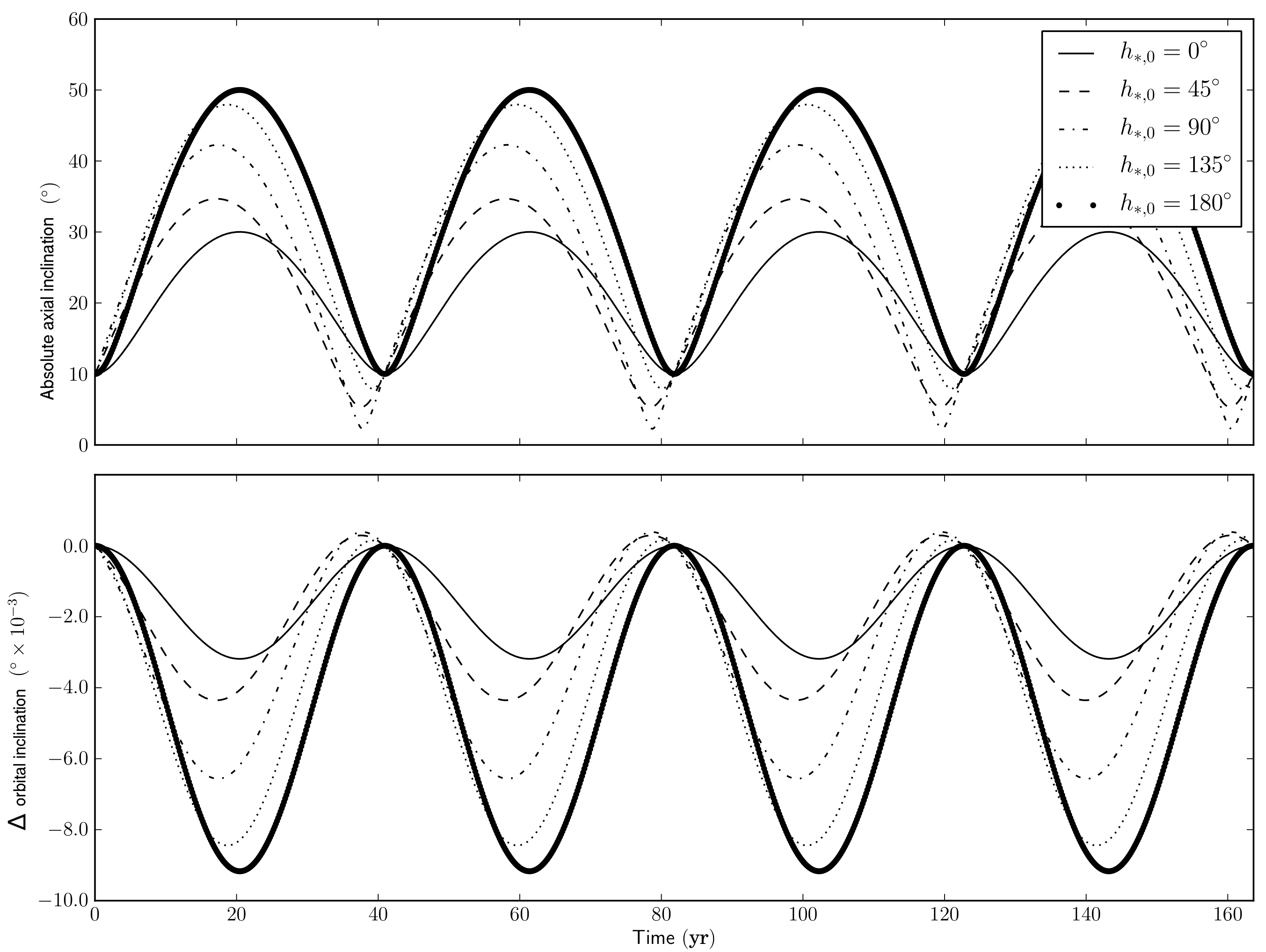}
\end{center}
\caption{Time evolution of the absolute axial (top) and orbital (bottom) inclinations
of a pulsar planetary two-body system. In the bottom panel the quantity
on the $y$ axis is the difference from the initial value of orbital
inclination. Time is measured in years.\label{fig:psr_plot}}
\end{figure*}

Figure \ref{fig:psr_plot} shows the time evolution of the absolute axial and
orbital inclinations of the planet for different initial values
of the angle $h_{\ast}$. The period of the evolution is much shorter
than in the previous case, amounting to $40$ years circa. This system
is lacking an almost-equilibrium point for $h_{\ast,0}=0$, as the
initial values of the inclinations are not close (although
the amplitude of the oscillation is still correlated to $h_{\ast,0}$).
As in the previous case, the amplitude of the oscillation for the
axial inclination dominates over the oscillation in orbital inclination, but since
the $\tilde{G}/G$ ratio is now larger the oscillation in orbital inclination
is also larger, reaching almost $0.01^{\circ}$ when $h_{\ast,0}=180^{\circ}$.

\subsection{Earth-orbiting gyroscope}
In this last example, we consider the gravitomagnetic effects on a
gyroscope in low-orbit on board of a spacecraft around the Earth.
The parameters of the system are taken from the experimental setup
of the Gravity Probe B mission \citep{PhysRevLett.106.221101}: the orbit is polar ($i=90.007^{\circ}$)
with a semi-major axis of $7027\:\mathrm{km}$ and low eccentricity
($e=0.0014$). The gyroscope consists of a rapidly rotating
quartz sphere ($38\:\mathrm{mm}$ diameter) whose spin axis is lying
on the Earth's equatorial plane (i.e., the spin plane is also ``polar'').
The spin vector of the gyroscope and the orbital angular momentum
vector of the spacecraft, both lying on the equatorial plane, are
perpendicular to each other. Translated into mean Delaunay and Serret-Andoyer
parameters, this initial geometric configuration implies $H\sim0$,
$\tilde{H}_{\ast}\sim0$, $G_{xy}\sim G$, $\tilde{G}_{xy\ast}\sim\tilde{G}$
and $h_{\ast}=\pm\pi/2$ (with the sign depending on the values of
$h$ and $\tilde{h}$ -- note that in \citet{PhysRevLett.106.221101} the configuration
shown in Fig. 1 implies $h_{\ast}=-\pi/2$). The substitution of these values
in the general formula for $H\left( t \right)$ yields a period of
roughly $195\:\mathrm{ka}$.

Turning now our attention to the equations of motion, we first note
that, given the timescale of the time-evolution for $H$, we can consider
$H$ and $h_{\ast}$ as constants for the duration of the Gravity
Probe B experiment. With this assumption, the equation of motion for
$\tilde{h}$ radically simplifies to
\begin{equation}
\frac{d\tilde{h}}{dt}=\frac{1}{2}\epsilon\frac{m_{2}^{3}\mathcal{G}^{4}J_{2}}{L^{3}G^{3}},
\end{equation}
and thus $\tilde{h}$ evolves linearly with time. This is the expected
frame-dragging precession, which, after substitution of the appropriate
numerical values, amounts to circa $0.04^{\prime\prime}$ per year
\citep{PhysRevLett.106.221101}. The effect corresponds to a movement
of the mean spin axis of the gyroscope along the equatorial plane in the
direction of the rotation of the Earth. The other expected relativistic
motion is the geodetic effect, which is a drift of the mean spin axis on
the orbital plane in the direction of the orbital motion. Translated
into mean Serret-Andoyer variables, this effect affects the element $\tilde{H}=\tilde{H}_{\ast}-H$.
Recalling that $\tilde{H}_{\ast}$ is a constant of motion and that
in our experimental setup $H\sim0$ and $G_{xy}\sim G$, the time
evolution of $\tilde{H}$ is then given by
\begin{equation}
\frac{d\tilde{H}}{dt}=-\frac{dH}{dt}=-\frac{3}{2}\epsilon\frac{\tilde{G}_{xy\ast}m_{2}^{4}\mathcal{G}^{4}}{L^{3}G^{2}}\sin h_{\ast}.
\end{equation}
Recalling now that $\tilde{H}=\tilde{G}\cos I$ by definition, where
$I$ is the inclination of the mean spin vector with respect to the $z$
axis and $\tilde{G}$ is a constant of motion, we can write
\begin{equation}
\frac{d\tilde{H}}{dt}=\tilde{G}\frac{d\cos I}{dt},
\end{equation}
and since $\tilde{G}\sim\tilde{G}_{xy\ast}$ and $I\sim\pi/2$, a Taylor
expansion of $\cos I$ gives the simplified formula
\begin{equation}
\frac{dI}{dt}=\frac{3}{2}\epsilon\frac{m_{2}^{4}\mathcal{G}^{4}}{L^{3}G^{2}}\sin h_{\ast}.
\end{equation}
Depending then on the orientation of the angular momentum and spin
vectors, $h_{\ast}=\pm\pi/2$ and the time variation of $I$ becomes
then
\begin{equation}
\frac{dI}{dt}=\pm\frac{3}{2}\epsilon\frac{m_{2}^{4}\mathcal{G}^{4}}{L^{3}G^{2}}.
\end{equation}
Substitution of the numerical values in this formula yields the expected
value of circa $\pm6.6{}^{\prime\prime}$ per year.
\section{Conclusions and future work}
In this paper we have analysed the post-Newtonian evolution of the restricted relativistic two-body problem with spin.
Our analysis has been performed via a modern perturbation scheme based on Lie series. The first important advantage of this method
is that it allows us to work out all the calculations in a closed analytic form, thus permitting
an analysis of long-term perturbative effects which are otherwise very difficult to detect with numerical techniques. Secondly, the Lie series formalism
can be iterated to higher post-Newtonian orders using essentially the same procedure employed in this study, and the connection between the transformed
coordinates at each perturbative step is given by explicit formul\ae{} (contrary to the classical von Zeipel method \citep{von_zeipel}).
Finally, the Lie series methodology can be easily implemented on computer algebra systems, thus relegating the most cumbersome parts of the development
of the theory to automatised algebraic manipulation.

Our analysis reproduces well-known classical results such as the relativistic pericentre precession, and the Lens-Thirring and geodetic effects.
In addition, we are able to thoroughly investigate the complex interplay between spins and orbit, and we
provide a novel solution for the full averaged equations of motion in terms of Weierstrass elliptic functions. This result
is particularly interesting as it establishes a connection with recent developments in the exact solution
of geodesic equations \citep[see, e.g.,][]{hackmann_analytical_2010,scharf_schwarzschild_2011,gibbons_application_2012},
which also employ elliptic and hyperelliptic functions.

From the mathematical point of view, our investigation highlights the existence of a rich dynamical environment, with multiple sets of fixed
points and aperiodic solutions arising with particular choices of initial conditions. From a physical point of view, we have shown
how some of these exotic configurations appear in correspondence of initial conditions for which
the approximation of our model starts to be unrealistic. In this sense, the generalisation of these results to a full two-body
model (to be tackled in an upcoming publication), will provide a more realistic insight into the physics of these particular solutions.

The application of our results to real physical systems shows how relativistic effects can accumulate over time to induce
substantial changes in the dynamics. In particular, the absolute orientation of the spin vector of planetary-sized bodies in the Solar System
undergoes significant changes, driven chiefly by the geodetic effect, over geological timescales.
In close pulsar planets, the evolution time shortens drastically to few years, and the
effects on the planet's orbit increase by multiple orders of magnitude. In this sense, our results could be used to devise new tests of General
Relativity based on precise measurements of the orbital parameters of pulsar planets and similar dynamical systems.

\bibliographystyle{mn2e}
\bibliography{biblio}

\begin{thebibliography}{}

\bibitem[\protect\citeauthoryear{Abramowitz \& Stegun}{Abramowitz \&
  Stegun}{1964}]{Abramowitz}
Abramowitz M.,  Stegun I.,  1964, Handbook of Mathematical Functions, fifth
  edn.
Dover, New York

\bibitem[\protect\citeauthoryear{Arnold}{Arnold}{1989}]{arnold_mathematical_1989}
Arnold V.~I.,  1989, Mathematical Methods of Classical Mechanics, 2nd edn.
Springer

\bibitem[\protect\citeauthoryear{Bailes, Bates, Bhalerao, Bhat, Burgay,
  Burke-Spolaor, D'Amico, Johnston, Keith, Kramer, Kulkarni, Levin, Lyne,
  Milia, Possenti, Spitler, Stappers \& van Straten}{Bailes
  et~al.}{2011}]{Bailes23092011}
Bailes M.,  Bates S.~D.,  Bhalerao V.,  Bhat N. D.~R.,  Burgay M.,
  Burke-Spolaor S.,  D'Amico N.,  Johnston S.,  Keith M.~J.,  Kramer M.,
  Kulkarni S.~R.,  Levin L.,  Lyne A.~G.,  Milia S.,  Possenti A.,  Spitler L.,
   Stappers B.,    van Straten W.,  2011, Science, 333, 1717

\bibitem[\protect\citeauthoryear{Barker \& {O'Connell}}{Barker \&
  {O'Connell}}{1970}]{Barker:1970vv}
Barker B.~M.,  {O'Connell} R.~F.,  1970, Physical Review D, 2, 1428

\bibitem[\protect\citeauthoryear{Barker \& {O'Connell}}{Barker \&
  {O'Connell}}{1975}]{barker_gravitational_1975}
Barker B.~M.,  {O'Connell} R.~F.,  1975, Physical Review D, 12, 329

\bibitem[\protect\citeauthoryear{Barker \& {O'Connell}}{Barker \&
  {O'Connell}}{1976}]{barker_lagrangian-hamiltonian_1976}
Barker B.~M.,  {O'Connell} R.~F.,  1976, Physical Review D, 14, 861

\bibitem[\protect\citeauthoryear{Barker \& {O'Connell}}{Barker \&
  {O'Connell}}{1979}]{BARKER:1979wb}
Barker B.~M.,  {O'Connell} R.~F.,  1979, General Relativity and Gravitation,
  11, 149

\bibitem[\protect\citeauthoryear{Biscani}{Biscani}{2008}]{biscani_design_2008}
Biscani F.,  2008, {Ph.D.} dissertation, University of Padua

\bibitem[\protect\citeauthoryear{Bogorodskii}{Bogorodskii}{1959}]{bogorodskii_relativistic_1959}
Bogorodskii A.~F.,  1959, Soviet Astronomy, 3, 857

\bibitem[\protect\citeauthoryear{Brumberg}{Brumberg}{1991}]{brumberg1991essential}
Brumberg V.,  1991, Essential relativistic celestial mechanics.
Taylor \& Francis

\bibitem[\protect\citeauthoryear{Cugusi \& Proverbio}{Cugusi \&
  Proverbio}{1978}]{cugusi_relativistic_1978}
Cugusi L.,  Proverbio E.,  1978, Astronomy and Astrophysics, 69, 321

\bibitem[\protect\citeauthoryear{Damour}{Damour}{2001}]{damour_coalescence_2001}
Damour T.,  2001, Physical Review D, 64

\bibitem[\protect\citeauthoryear{Damour, Jaranowski \& Sch\"{a}fer}{Damour
  et~al.}{2008}]{damour_effective_2008}
Damour T.,  Jaranowski P.,    Sch\"{a}fer G.,  2008, Physical Review D, 78,
  024009

\bibitem[\protect\citeauthoryear{Damour, Soffel \& Xu}{Damour
  et~al.}{1991a}]{1991PhRvD..43.3273D}
Damour T.,  Soffel M.,    Xu C.,  1991a, Physical Review D (Particles and
  Fields), 43, 3273

\bibitem[\protect\citeauthoryear{Damour, Soffel \& Xu}{Damour
  et~al.}{1993}]{1993PhRvD..47.3124D}
Damour T.,  Soffel M.,    Xu C.,  1993, Physical Review D (Particles and
  Fields), 47, 3124

\bibitem[\protect\citeauthoryear{Damour, Soffel \& Xu}{Damour
  et~al.}{1994}]{1994PhRvD..49..618D}
Damour T.,  Soffel M.,    Xu C.,  1994, Physical Review D (Particles and
  Fields), 49, 618

\bibitem[\protect\citeauthoryear{Damour, Soffel \& Xu}{Damour
  et~al.}{1991b}]{Damour:227658}
Damour T. M. A.~G.,  Soffel M.~H.,    Xu C.,  1991b, Phys. Rev. D, 45, 1017

\bibitem[\protect\citeauthoryear{de
  Sitter}{de~Sitter}{1916}]{de_sitter_einsteins_1916}
de Sitter W.,  1916, Monthly Notices of the Royal Astronomical Society, 77, 155

\bibitem[\protect\citeauthoryear{Deprit}{Deprit}{1969}]{deprit_canonical_1969}
Deprit A.,  1969, Celestial Mechanics, 1, 12

\bibitem[\protect\citeauthoryear{Deprit}{Deprit}{1982}]{deprit_delaunay_1982}
Deprit A.,  1982, Celestial Mechanics, 26, 9

\bibitem[\protect\citeauthoryear{Einstein}{Einstein}{1916}]{ANDP:ANDP19163540702}
Einstein A.,  1916, Annalen der Physik, 354, 769

\bibitem[\protect\citeauthoryear{Einstein, Infeld \& Hoffmann}{Einstein
  et~al.}{1938}]{Einstein:1938kq}
Einstein A.,  Infeld L.,    Hoffmann B.,  1938, The Annals of Mathematics, 39,
  65

\bibitem[\protect\citeauthoryear{Everitt et~al.,}{Everitt
  et~al.}{2011}]{PhysRevLett.106.221101}
Everitt C.,  et~al., 2011, Physical Review Letters, 106, 221101

\bibitem[\protect\citeauthoryear{Fock}{Fock}{1964}]{fock-book}
Fock V.~A.,  1964, The theory of space, time, and gravitation, 2d rev. ed.
  translated from the russian by n. kemmer. edn.
New York, Macmillan

\bibitem[\protect\citeauthoryear{Gibbons \& Vyska}{Gibbons \&
  Vyska}{2012}]{gibbons_application_2012}
Gibbons G.~W.,  Vyska M.,  2012, Classical and Quantum Gravity, 29, 065016

\bibitem[\protect\citeauthoryear{Gurfil, Elipe, Tangren \& Efroimsky}{Gurfil
  et~al.}{2007}]{gurfil_serret-andoyer_2007}
Gurfil P.,  Elipe A.,  Tangren W.,    Efroimsky M.,  2007, Regular and Chaotic
  Dynamics, 12, 389

\bibitem[\protect\citeauthoryear{Hackmann, L\"{a}mmerzahl, Kagramanova \&
  Kunz}{Hackmann et~al.}{2010}]{hackmann_analytical_2010}
Hackmann E.,  L\"{a}mmerzahl C.,  Kagramanova V.,    Kunz J.,  2010, Physical
  Review D, 81, 044020

\bibitem[\protect\citeauthoryear{Hees, Wolf, Lamine, Jaekel, Poncin-Lafitte,
  Lainey \& Dehant}{Hees et~al.}{2011}]{Hees:2011wt}
Hees A.,  Wolf P.,  Lamine B.,  Jaekel M.~T.,  Poncin-Lafitte C.~L.,  Lainey
  V.,    Dehant V.,  2011, {arXiv:1110.0659}

\bibitem[\protect\citeauthoryear{Heimberger, Soffel \& Ruder}{Heimberger
  et~al.}{1990}]{heimberger_relativistic_1990}
Heimberger J.,  Soffel M.,    Ruder H.,  1990, Celestial Mechanics and
  Dynamical Astronomy, 47, 205

\bibitem[\protect\citeauthoryear{Hori}{Hori}{1966}]{hori_theory_1966}
Hori G.,  1966, Publications of the Astronomical Society of Japan, 18

\bibitem[\protect\citeauthoryear{Johansson et~al.,}{Johansson
  et~al.}{2011}]{mpmath}
Johansson F.,  et~al., 2011, mpmath: a {P}ython library for arbitrary-precision
  floating-point arithmetic (version 0.17)

\bibitem[\protect\citeauthoryear{Kopeikin, Efroimsky \& Kaplan}{Kopeikin
  et~al.}{2011}]{kopeikin2011relativistic}
Kopeikin S.,  Efroimsky M.,    Kaplan G.,  2011, Relativistic Celestial
  Mechanics of the Solar System.
Wiley-VCH

\bibitem[\protect\citeauthoryear{Landau \& Lifschits}{Landau \&
  Lifschits}{1975}]{landau1975classical}
Landau L.,  Lifschits E.,  1975, The classical theory of fields.
Vol.~2, Butterworth-Heinemann

\bibitem[\protect\citeauthoryear{Morbidelli}{Morbidelli}{2002}]{morbidelli_modern_2002}
Morbidelli A.,  2002, Modern Celestial Mechanics: Dynamics in the Solar System,
  1st edn.
{CRC} Press

\bibitem[\protect\citeauthoryear{Morbidelli \& Giorgilli}{Morbidelli \&
  Giorgilli}{1993}]{morbidelli_quantitative_1993}
Morbidelli A.,  Giorgilli A.,  1993, Celestial Mechanics and Dynamical
  Astronomy, 55, 131

\bibitem[\protect\citeauthoryear{Murray \& Dermott}{Murray \&
  Dermott}{2000}]{murray_solar_2000}
Murray C.~D.,  Dermott S.~F.,  2000, Solar System Dynamics.
Cambridge University Press

\bibitem[\protect\citeauthoryear{Palaci\'{a}n}{Palaci\'{a}n}{2002}]{palacian_closed-form_2002}
Palaci\'{a}n J.,  2002, Chaos, Solitons \& Fractals, 13, 853

\bibitem[\protect\citeauthoryear{Papapetrou}{Papapetrou}{1974}]{papapetrou1974lectures}
Papapetrou A.,  1974, Lectures on general relativity.
Springer

\bibitem[\protect\citeauthoryear{Peebles}{Peebles}{1993}]{peebles1993principles}
Peebles P.,  1993, Principles of physical cosmology.
Princeton University Press

\bibitem[\protect\citeauthoryear{Richardson \& Kelly}{Richardson \&
  Kelly}{1988}]{1988CeMec..43..193R}
Richardson D.~L.,  Kelly T.~J.,  1988, Celestial Mechanics, 43, 193

\bibitem[\protect\citeauthoryear{Scharf}{Scharf}{2011}]{scharf_schwarzschild_2011}
Scharf G.,  2011, Journal of Modern Physics, 2, 274

\bibitem[\protect\citeauthoryear{Schiff}{Schiff}{1960a}]{schiff_motion_1960}
Schiff L.~I.,  1960a, Proceedings of the National Academy of Sciences of the
  United States of America, 46, 871

\bibitem[\protect\citeauthoryear{Schiff}{Schiff}{1960b}]{schiff1960experimental}
Schiff L.~I.,  1960b, American Journal of Physics, 28, 340

\bibitem[\protect\citeauthoryear{Straumann}{Straumann}{1984}]{straumann_general_1984}
Straumann N.,  1984, General relativity and relativistic astrophysics.
{Springer-Verlag}

\bibitem[\protect\citeauthoryear{Thirring}{Thirring}{1918}]{thirring_uber_1918}
Thirring H.,  1918, Physikalische Zeitschrift, 19, 33

\bibitem[\protect\citeauthoryear{von Zeipel}{von Zeipel}{1916}]{von_zeipel}
von Zeipel H.,  1916, {Arkiv f\"{o}r matematik, astronomi och fysik}, 11, 1

\bibitem[\protect\citeauthoryear{Wex}{Wex}{1995}]{wex_second_1995}
Wex N.,  1995, Classical and Quantum Gravity, 12, 983

\bibitem[\protect\citeauthoryear{Whittaker \& Watson}{Whittaker \&
  Watson}{1927}]{WhWa27}
Whittaker E.~T.,  Watson G.~N.,  1927, A Course of Modern Analysis, fourth edn.
Cambridge University Press

\bibitem[\protect\citeauthoryear{Wu \& Xie}{Wu \&
  Xie}{2010}]{wu_symplectic_2010}
Wu X.,  Xie Y.,  2010, Physical Review D, 81, 084045

\end{thebibliography}
\appendix
\section{Closed-form averaging}
\label{sec:closed_form_av}
We detail in this appendix the solution of eq. \eqref{eq:chi_integral}, here reproduced for convenience:
\begin{equation}
\chi=\int\frac{L^{3}}{\mathcal{G}^{2}m_{2}^{2}}\left(\mathcal{H}_{1}-\mathcal{K}\right)dl.
\label{eq:chi_integral_ag}
\end{equation}
From eq. \eqref{eq:Ham_F_1}, the structure of $\mathcal{H}_{1}$ is as follows:
\begin{align}
\mathcal{H}_{1} & =\mathcal{A}_{0}+\frac{\mathcal{A}_{1}}{r}+\frac{\mathcal{A}_{2}}{r^{2}}+\frac{1}{r^{3}}\left[\mathcal{A}_{3a}+\mathcal{A}_{3b}\cos\left(\tilde{h}-h\right)\right]\nonumber \\
 & \quad+\frac{1}{r^{3}}\left[\mathcal{B}_{0}\cos\left(2f+2g\right)+\mathcal{B}_{1}\cos\left(2f+2g+\tilde{h}-h\right)\right.\nonumber \\
 & \quad\left.+\mathcal{B}_{2}\cos\left(2f+2g-\tilde{h}+h\right)\right],\label{eq:H1_struct}
\end{align}
where the coefficients $\mathcal{A}_{i}$ and $\mathcal{B}_{i}$ are
functions of the momenta only. As previously remarked, in this expression we have to
consider $r$ and $f$ as implicit functions of $L$, $G$ and $l$. For the solution of
eq. \eqref{eq:chi_integral_ag}, it is convenient to look for a $\mathcal{K}$ in the form
\begin{equation}
\mathcal{K}=\mathcal{A}_{0}+\mathcal{K}_{1},
\label{eq:K_def}
\end{equation}
where $\mathcal{A}_{0}$ is a function of the momenta only from eq.
\eqref{eq:H1_struct}. Eq. \eqref{eq:chi_integral_ag} thus becomes:
\begin{align}
\chi %& =\frac{L^{3}}{\mathcal{G}^{2}m_{2}^{2}}\int\left(\mathcal{H}_{1}-\mathcal{A}_{0}-\mathcal{K}_{1}\right)dl\\
 & =\frac{L^{3}}{\mathcal{G}^{2}m_{2}^{2}}\left[\int\left(\mathcal{H}_{1}-\mathcal{A}_{0}\right)dl-\int\mathcal{K}_{1}dl\right].
\label{eq:chi_int_01}
\end{align}
We use now the differential relation \citep{murray_solar_2000}
\begin{equation}
dl=\frac{r^{2}}{a^{2}\sqrt{1-e^{2}}}df=\frac{r^{2}\mathcal{G}^{2}m_{2}^{2}}{GL^{3}}df
\end{equation}
to change the integration variable in the first integral of eq. \eqref{eq:chi_int_01}
from $l$ to $f$:
\begin{equation}
\chi=\frac{1}{G}\int r^{2}\left(\mathcal{H}_{1}-\mathcal{A}_{0}\right)df-\frac{L^{3}}{\mathcal{G}^{2}m_{2}^{2}}\int\mathcal{K}_{1}dl,\label{eq:chi_def}
\end{equation}
where the first integrand in eq. \eqref{eq:chi_def} has now become
\begin{align}
r^{2}\left(\mathcal{H}_{1}-\mathcal{A}_{0}\right) & =
r\mathcal{A}_{1}+\mathcal{A}_{2}+\frac{1}{r}\left[\mathcal{A}_{3a}+\mathcal{A}_{3b}\cos\left(\tilde{h}-h\right)\right]\nonumber \\
& \quad+\frac{1}{r}\left[\mathcal{B}_{0}\cos\left(2f+2g\right)
\vphantom{\mathcal{B}_{1}\cos\left(2f+2g+\tilde{h}-h\right)}
\right.\notag\\
& \quad+\mathcal{B}_{1}\cos\left(2f+2g+\tilde{h}-h\right)\nonumber \\
 & \quad\left.+\mathcal{B}_{2}\cos\left(2f+2g-\tilde{h}+h\right)\right],
\end{align}
and the coefficients $\mathcal{A}_{i}$ and $\mathcal{B}_{i}$ are still
functions of the momenta only. We now perform a further split of the first integrand
in eq. \eqref{eq:chi_def}, separating
the term $\mathcal{A}_1$ in its own integral:
\begin{align}
\chi & =\frac{1}{G}\int r\mathcal{A}_{1} df + \frac{1}{G} \int \left\{
\mathcal{A}_{2}+\frac{1}{r}\left[\mathcal{A}_{3a}+\mathcal{A}_{3b}\cos\left(\tilde{h}-h\right)\right]\right.\notag\\
&\quad +\frac{1}{r}\left[\mathcal{B}_{0}\cos\left(2f+2g\right)+\mathcal{B}_{1}\cos\left(2f+2g+\tilde{h}-h\right)\right.\nonumber \\
&\quad \left.+
\vphantom{\mathcal{A}_{2}+\frac{1}{r}\left[\mathcal{A}_{3a}+\mathcal{A}_{3b}\cos\left(\tilde{h}-h\right)\right]}
\left. \mathcal{B}_{2}\cos\left(2f+2g-\tilde{h}+h\right)\right]\right\} df -\frac{L^{3}}{\mathcal{G}^{2}m_{2}^{2}}\int\mathcal{K}_{1}dl.
\label{eq:chi_def_01}
\end{align}
The next step is to change the integration variable in the first integrand of eq. \eqref{eq:chi_def_01} from the true anomaly $f$
to the eccentric anomaly $E$ via the standard differential relation
\begin{equation}
df=\frac{a}{r}\sqrt{1-e^{2}}dE=\frac{LG}{r\mathcal{G}m_{2}}dE,
\end{equation}
where the eccentricity $e$ is to be considered an implicit function
of $L$ and $G$ and the eccentric anomaly $E$ an implicit function
of $L$, $G$ and $l$. Eq. \eqref{eq:chi_def_01} now reads:
\begin{align}
\chi & =\frac{1}{G}\int \frac{\mathcal{A}_{1}LG}{\mathcal{G}m_{2}} dE + \frac{1}{G} \int \left\{
\mathcal{A}_{2}+\frac{1}{r}\left[\mathcal{A}_{3a}+\mathcal{A}_{3b}\cos\left(\tilde{h}-h\right)\right]\right.\notag\\
&\quad +\frac{1}{r}\left[\mathcal{B}_{0}\cos\left(2f+2g\right)+\mathcal{B}_{1}\cos\left(2f+2g+\tilde{h}-h\right)\right.\nonumber \\
&\quad \left.+
\vphantom{\mathcal{A}_{2}+\frac{1}{r}\left[\mathcal{A}_{3a}+\mathcal{A}_{3b}\cos\left(\tilde{h}-h\right)\right]}
\left. \mathcal{B}_{2}\cos\left(2f+2g-\tilde{h}+h\right)\right]\right\} df -\frac{L^{3}}{\mathcal{G}^{2}m_{2}^{2}}\int\mathcal{K}_{1}dl.
\label{eq:chi_def_02}
\end{align}
By substituting the standard relation
\begin{equation}
\frac{1}{r} = \frac{1+e\cos f}{a\left(1-e^{2}\right)}=\frac{\mathcal{G}m_{2}}{G^{2}}+\frac{\mathcal{G}m_{2}}{G^{2}}e\cos f,
\end{equation}
in the second integrand of eq. \eqref{eq:chi_def_02}, and after straightforward algebraic passages, eq. \eqref{eq:chi_def_02} becomes
\begin{align}
\chi &
=\frac{1}{G}\left\{\int\mathcal{C}_{1}dE+\int\left[\mathcal{C}_{2,0}+\mathcal{C}_{2,1}\cos\left(\tilde{h}-h\right)\right]df\vphantom{\int\sum_{j\neq0,k,n,m}\mathcal{D}_{j,k,n,m}\left(jf+kg+n\tilde{h}+mh\right)df}\right.\nonumber \\
& \quad\left.+\int\sum_{j\neq0,k,n,m}\mathcal{D}_{jknm}\cos\left(jf+kg+n\tilde{h}+mh\right)df\right\}\notag\\
&\quad -\frac{L^{3}}{\mathcal{G}^{2}m_{2}^{2}}\int\mathcal{K}_{1}dl,
\end{align}
where the $\mathcal{C}$ and $\mathcal{D}$ coefficients are functions of the momenta only,
and the true anomaly $f$ always appears in the cosines associated to the $\mathcal{D}$ coefficients.
We can now choose $\mathcal{K}_{1}$ as
\begin{equation}
\mathcal{K}_{1}=\frac{\mathcal{G}^{2}m_{2}^{2}}{GL^{3}}\left[\mathcal{C}_{1}+\mathcal{C}_{2,0}+\mathcal{C}_{2,1}\cos\left(\tilde{h}-h\right)\right],
\label{eq:K1_def}
\end{equation}
so that the final expression for the generator $\chi$ becomes
\begin{align}
\chi & =\frac{1}{G}\left\{\mathcal{C}_{1}\left(E-l\right)+\left[\mathcal{C}_{2,0}
+\mathcal{C}_{2,1}\cos\left(\tilde{h}-h\right)\right]\left(f-l\right)
\vphantom{\frac{1}{G}\int\sum_{j\neq0,k,n,m}\mathcal{D}_{jknm}\cos\left(jf+kg+n\tilde{h}+mh\right)df}
\right.\nonumber \\
&\left.\quad+\int\sum_{j\neq0,k,n,m}\mathcal{D}_{jknm}\cos\left(jf+kg+n\tilde{h}+mh\right)df\right\},
\end{align}
where the integral in this expression is trivially calculated (as the $\mathcal{D}$ coefficients do not depend upon $f$).
$\chi$ has thus been defined as a $2\pi$-periodic function in the angular coordinates. By plugging eq. \eqref{eq:K1_def}
into eq. \eqref{eq:K_def}, we can now see how
\begin{equation}
\mathcal{K}=\mathcal{E}_{0}+\mathcal{E}_{1}\cos\left(\tilde{h}-h\right),
\end{equation}
where the $\mathcal{E}$ coefficients are functions of the momenta
only. Finally, via this definition of $\mathcal{K}$ and eqs. \eqref{eq:h_eq_00} and \eqref{eq:lie_S_prime}, we obtain eq. \eqref{eq:H_av_EE}:
\begin{equation}
\mathcal{H}^{\prime}=\mathcal{H}_\textnormal{N}+\epsilon\left[\mathcal{E}_{0}+\mathcal{E}_{1}\cos\left(\tilde{h}-h\right)\right].
\end{equation}
\section{Partial derivatives of $\mathcal{F}_0$ and $\mathcal{F}_1$}
\label{sec:full_form}
The full form of the partial derivatives of $\mathcal{F}_0$ with respect to the mean momenta appearing in eqs. \eqref{eq:averaged_angles} is:
\begin{align}
\frac{\partial \mathcal{F}_0}{\partial L} & = -\frac{3}{2}\frac{{J_2}{\mathcal{G}}^{4}{\tilde{H}_\ast}{m_2}^{3}}{{G}^{3}{L}^{4}}
-\frac{15}{2}\frac{{\mathcal{G}}^{4}{m_2}^{4}}{{L}^{5}}
-\frac{9}{2}\frac{{H}{\mathcal{G}}^{4}{\tilde{H}_\ast}{m_2}^{4}}{{G}^{3}{L}^{4}}\notag\\
&\quad +\frac{9}{2}\frac{{H}^{2}{\mathcal{G}}^{4}{m_2}^{4}}{{G}^{3}{L}^{4}}
-\frac{9}{2}\frac{{H}^{3}{J_2}{\mathcal{G}}^{4}{m_2}^{3}}{{G}^{5}{L}^{4}}
+\frac{9}{2}\frac{{H}^{2}{J_2}{\mathcal{G}}^{4}{\tilde{H}_\ast}{m_2}^{3}}{{G}^{5}{L}^{4}}\notag\\
&\quad +9\frac{{\mathcal{G}}^{4}{m_2}^{4}}{{G}{L}^{4}}-\frac{9}{2}\frac{{H}{J_2}{\mathcal{G}}^{4}{m_2}^{3}}{{G}^{3}{L}^{4}},\\
\frac{\partial \mathcal{F}_0}{\partial G} & = -\frac{9}{2}\frac{{H}{J_2}{\mathcal{G}}^{4}{m_2}^{3}}{{G}^{4}{L}^{3}}
+\frac{9}{2}\frac{{H}^{2}{\mathcal{G}}^{4}{m_2}^{4}}{{G}^{4}{L}^{3}}
+\frac{15}{2}\frac{{H}^{2}{J_2}{\mathcal{G}}^{4}{\tilde{H}_\ast}{m_2}^{3}}{{G}^{6}{L}^{3}}\notag\\
&\quad +3\frac{{\mathcal{G}}^{4}{m_2}^{4}}{{G}^{2}{L}^{3}}
-\frac{3}{2}\frac{{J_2}{\mathcal{G}}^{4}{\tilde{H}_\ast}{m_2}^{3}}{{G}^{4}{L}^{3}}
-\frac{15}{2}\frac{{H}^{3}{J_2}{\mathcal{G}}^{4}{m_2}^{3}}{{G}^{6}{L}^{3}}\notag\\
&\quad -\frac{9}{2}\frac{{H}{\mathcal{G}}^{4}{\tilde{H}_\ast}{m_2}^{4}}{{G}^{4}{L}^{3}},\\
\frac{\partial \mathcal{F}_0}{\partial H} & = \frac{3}{2}\frac{{J_2}{\mathcal{G}}^{4}{m_2}^{3}}{{G}^{3}{L}^{3}}
-3\frac{{H}{J_2}{\mathcal{G}}^{4}{\tilde{H}_\ast}{m_2}^{3}}{{G}^{5}{L}^{3}}
+\frac{9}{2}\frac{{H}^{2}{J_2}{\mathcal{G}}^{4}{m_2}^{3}}{{G}^{5}{L}^{3}}\notag\\
&\quad -3\frac{{H}{\mathcal{G}}^{4}{m_2}^{4}}{{G}^{3}{L}^{3}}+\frac{3}{2}\frac{{\mathcal{G}}^{4}{\tilde{H}_\ast}{m_2}^{4}}{{G}^{3}{L}^{3}},\\
\frac{\partial \mathcal{F}_0}{\partial \tilde{H}_\ast} & = \frac{1}{2}\frac{{J_2}{\mathcal{G}}^{4}{m_2}^{3}}{{G}^{3}{L}^{3}}+\frac{3}{2}\frac{{H}{\mathcal{G}}^{4}{m_2}^{4}}{{G}^{3}{L}^{3}}-\frac{3}{2}\frac{{H}^{2}{J_2}{\mathcal{G}}^{4}{m_2}^{3}}{{G}^{5}{L}^{3}}.
\end{align}
The partial derivatives of $\mathcal{F}_1$ are instead:
\begin{align}
\frac{\partial \mathcal{F}_1}{\partial L} & = -\frac{9}{2}\frac{{G_{xy}}{\mathcal{G}}^{4}{\tilde{G}_{xy\ast}}{m_2}^{4}}{{G}^{3}{L}^{4}}+\frac{9}{2}\frac{{G_{xy}}{H}{J_2}{\mathcal{G}}^{4}{\tilde{G}_{xy\ast}}{m_2}^{3}}{{G}^{5}{L}^{4}},\\
\frac{\partial \mathcal{F}_1}{\partial G} & = \frac{15}{2}\frac{{G_{xy}}{H}{J_2}{\mathcal{G}}^{4}{\tilde{G}_{xy\ast}}{m_2}^{3}}{{G}^{6}{L}^{3}}
-\frac{9}{2}\frac{{G_{xy}}{\mathcal{G}}^{4}{\tilde{G}_{xy\ast}}{m_2}^{4}}{{G}^{4}{L}^{3}}\notag\\
&\quad +\frac{3}{2}\frac{{\mathcal{G}}^{4}{\tilde{G}_{xy\ast}}{m_2}^{4}}{{G}^{2}{G_{xy}}{L}^{3}}
-\frac{3}{2}\frac{{H}{J_2}{\mathcal{G}}^{4}{\tilde{G}_{xy\ast}}{m_2}^{3}}{{G}^{4}{G_{xy}}{L}^{3}},\\
\frac{\partial \mathcal{F}_1}{\partial H} & = \frac{3}{2}\frac{{G_{xy}}{H}^{2}{J_2}{\mathcal{G}}^{4}{m_2}^{3}}{{G}^{5}{L}^{3}{\tilde{G}_{xy\ast}}}
+\frac{3}{2}\frac{{H}^{2}{J_2}{\mathcal{G}}^{4}{\tilde{G}_{xy\ast}}{m_2}^{3}}{{G}^{5}{G_{xy}}{L}^{3}}\notag\\
&\quad +\frac{3}{2}\frac{{G_{xy}}{\mathcal{G}}^{4}{\tilde{H}_\ast}{m_2}^{4}}{{G}^{3}{L}^{3}{\tilde{G}_{xy\ast}}}
-\frac{3}{2}\frac{{G_{xy}}{J_2}{\mathcal{G}}^{4}{\tilde{G}_{xy\ast}}{m_2}^{3}}{{G}^{5}{L}^{3}}\notag\\
&\quad -\frac{3}{2}\frac{{G_{xy}}{H}{\mathcal{G}}^{4}{m_2}^{4}}{{G}^{3}{L}^{3}{\tilde{G}_{xy\ast}}}
-\frac{3}{2}\frac{{H}{\mathcal{G}}^{4}{\tilde{G}_{xy\ast}}{m_2}^{4}}{{G}^{3}{G_{xy}}{L}^{3}}\notag\\
&\quad -\frac{3}{2}\frac{{G_{xy}}{H}{J_2}{\mathcal{G}}^{4}{\tilde{H}_\ast}{m_2}^{3}}{{G}^{5}{L}^{3}{\tilde{G}_{xy\ast}}},\label{eq:F1_H}\\
\frac{\partial \mathcal{F}_1}{\partial \tilde{G}} & = \frac{3}{2}\frac{{G_{xy}}{\mathcal{G}}^{4}{\tilde{G}}{m_2}^{4}}{{G}^{3}{L}^{3}{\tilde{G}_{xy\ast}}}-\frac{3}{2}\frac{{G_{xy}}{H}{J_2}{\mathcal{G}}^{4}{\tilde{G}}{m_2}^{3}}{{G}^{5}{L}^{3}{\tilde{G}_{xy\ast}}},\label{eq:F1_Gt}\\
\frac{\partial \mathcal{F}_1}{\partial \tilde{H}_\ast} & = -\frac{3}{2}\frac{{G_{xy}}{\mathcal{G}}^{4}{\tilde{H}_\ast}{m_2}^{4}}{{G}^{3}{L}^{3}{\tilde{G}_{xy\ast}}}
-\frac{3}{2}\frac{{G_{xy}}{H}^{2}{J_2}{\mathcal{G}}^{4}{m_2}^{3}}{{G}^{5}{L}^{3}{\tilde{G}_{xy\ast}}}\notag\\
&\quad +\frac{3}{2}\frac{{G_{xy}}{H}{\mathcal{G}}^{4}{m_2}^{4}}{{G}^{3}{L}^{3}{\tilde{G}_{xy\ast}}}
+\frac{3}{2}\frac{{G_{xy}}{H}{J_2}{\mathcal{G}}^{4}{\tilde{H}_\ast}{m_2}^{3}}{{G}^{5}{L}^{3}{\tilde{G}_{xy\ast}}}.\label{eq:F1_Hts}
\end{align}
\section{Coefficients of the quartic polynomial}
\label{sec:cf_f4H}
Here follows the full form of the coefficients of the polynomial $f_4\left(H\right)$ in eq. \eqref{eq:f4H_def}:
\begin{align}
a_0 & = -\frac{9}{4}\frac{{J_2}^{2}{\mathcal{G}}^{8}{\tilde{G}}^{2}{\epsilon}^{2}{m_2}^{6}}{{G}^{10}{L}^{6}}
-\frac{27}{4}\frac{{J_2}^{2}{\mathcal{G}}^{8}{\epsilon}^{2}{m_2}^{6}}{{G}^{8}{L}^{6}},\\
a_1 & = \frac{9}{8}\frac{{J_2}{\mathcal{G}}^{8}{\tilde{G}}^{2}{\epsilon}^{2}{m_2}^{7}}{{G}^{8}{L}^{6}}
+\frac{15}{8}\frac{{J_2}^{2}{\mathcal{G}}^{8}{\tilde{H}_\ast}{\epsilon}^{2}{m_2}^{6}}{{G}^{8}{L}^{6}}
-\frac{45}{32}\frac{{J_2}{\mathcal{G}}^{8}{\epsilon}^{2}{m_2}^{7}}{{G}^{5}{L}^{7}}\notag\\
&\quad+\frac{9}{2}\frac{{J_2}{\mathcal{G}}^{8}{\epsilon}^{2}{m_2}^{7}}{{G}^{6}{L}^{6}}
-\frac{3}{8}\frac{{J_2}{\mathcal{G}}^{4}{\mathcal{I}_1}{\tilde{G}}^{2}{\epsilon}{m_2}^{3}}{{G}^{5}{L}^{3}}
+\frac{3}{8}\frac{{J_2}{\mathcal{G}}^{6}{\epsilon}{m_2}^{5}}{{G}^{5}{L}^{5}}\notag\\
&\quad+\frac{3}{4}\frac{{J_2}{\mathcal{G}}^{4}{\mathcal{H}^\prime}{\epsilon}{m_2}^{3}}{{G}^{5}{L}^{3}},\\
a_2 & = \frac{1}{4}\frac{{\mathcal{G}}^{4}{\mathcal{I}_1}{\tilde{G}}^{2}{\epsilon}{m_2}^{4}}{{G}^{3}{L}^{3}}
-\frac{3}{8}\frac{{J_2}^{2}{\mathcal{G}}^{8}{\epsilon}^{2}{m_2}^{6}}{{G}^{6}{L}^{6}}
-\frac{1}{8}\frac{{J_2}^{2}{\mathcal{G}}^{8}{\tilde{H}_\ast}^{2}{\epsilon}^{2}{m_2}^{6}}{{G}^{8}{L}^{6}}\notag\\
&\quad+\frac{15}{16}\frac{{\mathcal{G}}^{8}{\epsilon}^{2}{m_2}^{8}}{{G}^{3}{L}^{7}}
-\frac{1}{2}\frac{{\mathcal{G}}^{4}{\mathcal{H}^\prime}{\epsilon}{m_2}^{4}}{{G}^{3}{L}^{3}}
-\frac{1}{4}\frac{{\mathcal{G}}^{6}{\epsilon}{m_2}^{6}}{{G}^{3}{L}^{5}}\notag\\
&\quad+\frac{15}{16}\frac{{J_2}{\mathcal{G}}^{8}{\tilde{H}_\ast}{\epsilon}^{2}{m_2}^{7}}{{G}^{5}{L}^{7}}
+\frac{3}{8}\frac{{J_2}^{2}{\mathcal{G}}^{8}{\tilde{G}}^{2}{\epsilon}^{2}{m_2}^{6}}{{G}^{8}{L}^{6}}
-\frac{3}{8}\frac{{\mathcal{G}}^{8}{\tilde{G}}^{2}{\epsilon}^{2}{m_2}^{8}}{{G}^{6}{L}^{6}}\notag\\
&\quad-\frac{15}{8}\frac{{\mathcal{G}}^{8}{\epsilon}^{2}{m_2}^{8}}{{G}^{4}{L}^{6}}
+\frac{1}{4}\frac{{J_2}{\mathcal{G}}^{4}{\mathcal{I}_1}{\tilde{G}}^{2}{\tilde{H}_\ast}{\epsilon}{m_2}^{3}}{{G}^{5}{L}^{3}}\notag\\
&\quad-\frac{1}{2}\frac{{J_2}{\mathcal{G}}^{4}{\mathcal{H}^\prime}{\tilde{H}_\ast}{\epsilon}{m_2}^{3}}{{G}^{5}{L}^{3}}
-\frac{1}{4}\frac{{J_2}{\mathcal{G}}^{6}{\tilde{H}_\ast}{\epsilon}{m_2}^{5}}{{G}^{5}{L}^{5}}\notag\\
&\quad-\frac{7}{2}\frac{{J_2}{\mathcal{G}}^{8}{\tilde{H}_\ast}{\epsilon}^{2}{m_2}^{7}}{{G}^{6}{L}^{6}},\\
a_3 & = -\frac{3}{8}\frac{{J_2}{\mathcal{G}}^{4}{\mathcal{I}_1}{\tilde{G}}^{2}{\epsilon}{m_2}^{3}}{{G}^{3}{L}^{3}}
+\frac{3}{4}\frac{{J_2}{\mathcal{G}}^{4}{\mathcal{H}^\prime}{\epsilon}{m_2}^{3}}{{G}^{3}{L}^{3}}
-\frac{45}{32}\frac{{\mathcal{G}}^{8}{\tilde{H}_\ast}{\epsilon}^{2}{m_2}^{8}}{{G}^{3}{L}^{7}}\notag\\
&\quad+\frac{3}{4}\frac{{\mathcal{G}}^{4}{\mathcal{H}^\prime}{\tilde{H}_\ast}{\epsilon}{m_2}^{4}}{{G}^{3}{L}^{3}}
+\frac{3}{4}\frac{{J_2}{\mathcal{G}}^{8}{\tilde{H}_\ast}^{2}{\epsilon}^{2}{m_2}^{7}}{{G}^{6}{L}^{6}}\notag\\
&\quad+\frac{3}{8}\frac{{J_2}{\mathcal{G}}^{6}{\epsilon}{m_2}^{5}}{{G}^{3}{L}^{5}}
-\frac{3}{8}\frac{{J_2}^{2}{\mathcal{G}}^{8}{\tilde{H}_\ast}{\epsilon}^{2}{m_2}^{6}}{{G}^{6}{L}^{6}}
+\frac{3}{8}\frac{{\mathcal{G}}^{6}{\tilde{H}_\ast}{\epsilon}{m_2}^{6}}{{G}^{3}{L}^{5}}\notag\\
&\quad+\frac{9}{4}\frac{{J_2}{\mathcal{G}}^{8}{\epsilon}^{2}{m_2}^{7}}{{G}^{4}{L}^{6}}
-\frac{9}{8}\frac{{J_2}{\mathcal{G}}^{8}{\tilde{G}}^{2}{\epsilon}^{2}{m_2}^{7}}{{G}^{6}{L}^{6}}
+\frac{27}{8}\frac{{\mathcal{G}}^{8}{\tilde{H}_\ast}{\epsilon}^{2}{m_2}^{8}}{{G}^{4}{L}^{6}}\notag\\
&\quad-\frac{45}{32}\frac{{J_2}{\mathcal{G}}^{8}{\epsilon}^{2}{m_2}^{7}}{{G}^{3}{L}^{7}}
-\frac{3}{8}\frac{{\mathcal{G}}^{4}{\mathcal{I}_1}{\tilde{G}}^{2}{\tilde{H}_\ast}{\epsilon}{m_2}^{4}}{{G}^{3}{L}^{3}},\\
a_4 & = -\frac{1}{2}\frac{{J_2}{\mathcal{G}}^{4}{\mathcal{I}_1}{\tilde{G}}^{2}{\tilde{H}_\ast}{\epsilon}{m_2}^{3}}{{G}^{3}{L}^{3}}
+\frac{1}{2}\frac{{J_2}{\mathcal{G}}^{6}{\tilde{H}_\ast}{\epsilon}{m_2}^{5}}{{G}^{3}{L}^{5}}\notag\\
&\quad+\frac{15}{8}\frac{{\mathcal{G}}^{6}{\epsilon}{m_2}^{6}}{{L}^{6}}+{\mathcal{H}^\prime}{\mathcal{I}_1}{\tilde{G}}^{2}
+\frac{1}{2}\frac{{\mathcal{G}}^{2}{\mathcal{I}_1}{\tilde{G}}^{2}{m_2}^{2}}{{L}^{2}}\notag\\
&\quad-\frac{1}{4}\frac{{\mathcal{G}}^{4}{m_2}^{4}}{{L}^{4}}
-\frac{{\mathcal{G}}^{2}{\mathcal{H}^\prime}{m_2}^{2}}{{L}^{2}}
-\frac{15}{8}\frac{{\mathcal{G}}^{4}{\mathcal{I}_1}{\tilde{G}}^{2}{\epsilon}{m_2}^{4}}{{L}^{4}}\notag\\
&\quad+\frac{{J_2}{\mathcal{G}}^{4}{\mathcal{H}^\prime}{\tilde{H}_\ast}{\epsilon}{m_2}^{3}}{{G}^{3}{L}^{3}}
-{\mathcal{H}^\prime}^{2}+\frac{15}{4}\frac{{\mathcal{G}}^{4}{\mathcal{H}^\prime}{\epsilon}{m_2}^{4}}{{L}^{4}}
-\frac{1}{4}{\mathcal{I}_1}^{2}{\tilde{G}}^{4}\notag\\
&\quad-\frac{1}{4}\frac{{J_2}^{2}{\mathcal{G}}^{8}{\tilde{H}_\ast}^{2}{\epsilon}^{2}{m_2}^{6}}{{G}^{6}{L}^{6}}
-9\frac{{\mathcal{G}}^{8}{\epsilon}^{2}{m_2}^{8}}{{G}^{2}{L}^{6}}
+3\frac{{\mathcal{G}}^{4}{\mathcal{I}_1}{\tilde{G}}^{2}{\epsilon}{m_2}^{4}}{{G}{L}^{3}}\notag\\
&\quad-\frac{225}{64}\frac{{\mathcal{G}}^{8}{\epsilon}^{2}{m_2}^{8}}{{L}^{8}}
+\frac{45}{4}\frac{{\mathcal{G}}^{8}{\epsilon}^{2}{m_2}^{8}}{{G}{L}^{7}}
+\frac{9}{4}\frac{{\mathcal{G}}^{8}{\tilde{G}}^{2}{\epsilon}^{2}{m_2}^{8}}{{G}^{4}{L}^{6}}\notag\\
&\quad-3\frac{{\mathcal{G}}^{6}{\epsilon}{m_2}^{6}}{{G}{L}^{5}}
-\frac{9}{4}\frac{{\mathcal{G}}^{8}{\tilde{H}_\ast}^{2}{\epsilon}^{2}{m_2}^{8}}{{G}^{4}{L}^{6}}
+3\frac{{J_2}{\mathcal{G}}^{8}{\tilde{H}_\ast}{\epsilon}^{2}{m_2}^{7}}{{G}^{4}{L}^{6}}\notag\\
&\quad-\frac{15}{8}\frac{{J_2}{\mathcal{G}}^{8}{\tilde{H}_\ast}{\epsilon}^{2}{m_2}^{7}}{{G}^{3}{L}^{7}}
-6\frac{{\mathcal{G}}^{4}{\mathcal{H}^\prime}{\epsilon}{m_2}^{4}}{{G}{L}^{3}}.
\end{align}
\section{Removal of the sign ambiguity in the explicit solution for $H$}
\label{sec:sign_removal}
In this appendix we detail the removal of the sign ambiguity in eq. \eqref{eq:ell_inv_H}, here reproduced for convenience:
\begin{equation}
\int_{H_{0}}^{H}\pm\frac{dx}{\sqrt{f_{4}(x)}}=\int_{t_{0}}^{t}d\tau.
\label{eq:W_int_again}
\end{equation}
We first note how the sign change happens in correspondence of the
zeroes of the polynomial $f_{4}\left(x\right)$. Secondly, we note
how Weierstrass's elliptic function $\wp\left(z\right)$ is an even function,
whereas $\wp^{\prime}\left(z\right)$
is odd. Finally, from eq. \eqref{eq:weier_inv}, we introduce the shorthand
\begin{align}
\mathcal{W}\left(a;z\right) & =
\frac{1}{2\left[\wp\left(z\right)-\frac{1}{24}f_{4}^{\prime\prime}\left(a\right)\right]^{2}-\frac{1}{48}f_{4}\left(a\right)f_{4}^{iv}\left(a\right)}\notag\\
&\quad\cdot\left\{\frac{1}{2}f_{4}^{\prime}\left(a\right)\left[\wp\left(z\right)-\frac{1}{24}f_{4}^{\prime\prime}\left(a\right)\right]\right.\notag\\
&\quad\left.+\frac{1}{24}f_{4}\left(a\right)f_{4}^{\prime\prime\prime}\left(a\right)+\sqrt{f_{4}\left(a\right)}\wp^{\prime}\left(z\right)\right\},
\end{align}
in order to simplify the notation. Without loss of generality, we
suppose that the initial values of $h_{\ast}$ and $H$ are such that
the initial value of the time derivative of $H$ is positive and the
plus sign is then chosen in the integrand. We can integrate from $H_{0}$
to the next sign change in correspondence of a zero $H_{1}=H\left(t_{1}\right)$
of $f_{4}\left(H\right)$ and obtain
\begin{equation}
H_{1}=H_{0}+\mathcal{W}\left(H_{0};t_{1}-t_{0}\right).
\end{equation}
Next, taking $H_{1}$ as new initial value, and $H_{2}=H\left(t_{2}\right)$
as the next zero of $f_{4}\left(H\right)$, we can write
\begin{equation}
H_{2}=H_{1}+\mathcal{W}\left(H_{1};\pm t_{1}\mp t_{2}\right),
\end{equation}
where the plus-minus signs take into account the possibility of sign
change of the integrand. However, since $H_{1}$ is now a zero of
$f_{4}\left(H\right)$, the $\mathcal{W}$ function simplifies to
\begin{equation}
\mathcal{W}\left(H_{1};t_{1}-t_{2}\right)=\frac{1}{4}\frac{f_{4}^{\prime}\left(H_{1}\right)}{\wp\left(t_{1}-t_{2}\right)-\frac{1}{24}f_{4}^{\prime\prime}\left(H_{1}\right)},
\end{equation}
thus becoming an even function and allowing us to choose either sign
freely:
\begin{equation}
H_{2}=H_{1}+\mathcal{W}\left(H_{1};t_{2}-t_{1}\right).
\end{equation}
We can repeat this process as many times as needed, and write for
a generic value $H$
\begin{align}
H & =H_{0}+\mathcal{W}\left(H_{0};t_{1}-t_{0}\right)+\sum_{i=1}^{n-1}\mathcal{W}\left(H_{i};t_{i+1}-t_{i}\right)\notag\\
&\quad+\mathcal{W}\left(H_{n};t-t_{n}\right),
\end{align}
where $H_{1},\ldots,H_{n}$ are zeroes of $f_{4}\left(H\right)$.
This expression is the same we would have obtained if we had just
fixed in eq. \eqref{eq:W_int_again} the sign of the integrand once
and for all based on the initial signs of $\sin h_{\ast}$ and $\mathcal{F}_{1}$
and repeated the same procedure. Analogously, when the initial sign
is negative we can write
\begin{align}
H & =H_{0}+\mathcal{W}\left(H_{0};t_{0}-t_{1}\right)+\sum_{i=1}^{n-1}\mathcal{W}\left(H_{i};t_{i}-t_{i+1}\right)\notag\\
&\quad+\mathcal{W}\left(H_{n};t_{n}-t\right).
\end{align}
It follows then that the sign in eq. \eqref{eq:W_int_again} can be selected according to the initial signs of
$\sin h_{\ast}$ and $\mathcal{F}_{1}$, and the time evolution of $H$ is thus given by eq. \eqref{eq:H_of_t}.

\bsp

\label{lastpage}

\end{document}